\documentclass[fleqn,usenatbib]{mnras}

\usepackage{mathptmx}

\usepackage[T1]{fontenc}

\DeclareRobustCommand{\VAN}[3]{#2}
\let\VANthebibliography\thebibliography
\def\thebibliography{\DeclareRobustCommand{\VAN}[3]{##3}\VANthebibliography}

\usepackage{graphicx}	
\usepackage{amsmath}	
\usepackage{subfig}
\usepackage{amsmath,siunitx} 
\usepackage{lineno}
\usepackage{xcolor}

\newcommand{\hii}{H\textsc{ii}}
\newcommand{\mjy}{mJy beam$^{-1}$}
\newcommand{\ujy}{\textmu Jy beam$^{-1}$}
\newcommand{\s}{$\sim$}

\title[SARAO MeerKAT GPS Compact Sources]{The SARAO MeerKAT Galactic Plane Survey compact source catalogue}

\author[M. Mutale et al.]
{M. Mutale,$^{1}$\thanks{E-mail: M.Mutale@leeds.ac.uk (MM)}
M. A. Thompson,$^{1,2}$\thanks{E-mail: M.A.Thompson@leeds.ac.uk (MAT)}
G. M. Williams,$^{1,3}$
A. J. Rigby,$^{1}$
M. G. Hoare,$^{1}$
J. S. Urquhart,$^{4}$
\newauthor
M. F. Bietenholz,$^{5,6}$
C. Bordiu,$^{2}$
F. Camilo,$^{7}$
W. D. Cotton,$^{7,8}$
S. Goedhart,$^{7,9}$
W. O. Obonyo$^{10,11}$
\newauthor
S. Riggi,$^{2}$
A. Y. Yang$^{12,13}$\\
$^{1}$School of Physics and Astronomy, University of Leeds, LS2 9JT, UK\\
$^{2}$INAF-Osservatorio Astrofisico di Catania, Via Santa Sofia 78, 95123 Catania, Italy\\
$^{3}$Department of Physics, Aberystwyth University, Ceredigion, Cymru, SY23 3BZ, UK\\
$^{4}$Centre for Astrophysics and Planetary Science, University of Kent, Canterbury, CT2 7NH, UK\\
$^{5}$SARAO/Hartebeesthoek Radio Astronomy Observatory, PO Box 443, Krugersdorp, 1740, South Africa\\
$^{6}$Department of Physics and Astronomy, York University, Toronto, M3J 1P3, Ontario, Canada\\
$^{7}$South African Radio Astronomy Observatory, 2 Fir Street, Observatory, 7925, South Africa\\
$^{8}$National Radio Astronomy Observatory, 520 Edgemont Road, Char-
lottesville, VA 22903, USA\\
$^{9}$SKA Observatory, 2 Fir Street, Observatory 7925, South Africa\\
$^{10}$Department of Mathematical Sciences, University of South Africa, Cnr Christian de Wet Rd and Pioneer Avenue, Florida Park, 1709 Roodepoort, South Africa\\
$^{11}$Department of Astronomy and Space Science, The Technical University of Kenya, P.O. Box 52428 - 00200 Nairobi, Kenya\\
$^{12}$National Astronomical Observatories, Chinese Academy of Sciences, Beijing 100101, China\\
$^{13}$Key Laboratory of Radio Astronomy and Technology, Chinese Academy of Sciences, A20 Datun Road, Chaoyang District, Beijing, 100101, China
}

\date{Accepted XXX. Received YYY; in original form ZZZ}

\pubyear{2025}

\begin{document}
\label{firstpage}
\pagerange{\pageref{firstpage}--\pageref{lastpage}}
\maketitle

\begin{abstract}
We present a catalogue of compact sources detected in the SARAO MeerKAT 1.3 GHz Galactic Plane Survey (SMGPS). We extract 510\,599 compact sources, with areas less than five 8\arcsec\ beams, from the survey maps covering the regions $\ang{252} < l < \ang{358}$ and $\ang{2} < l < \ang{61}$ at $|b| \leq 1.5\degr$, which have an angular resolution of 8\arcsec\ and a sensitivity of $\sim$ 10-30 \textmu Jy beam$^{-1}$. In this paper, we describe the source identification and characterisation methods, present the quality assurance of the catalogue, explore the nature of the catalogue sources, and present initial science highlights. We limit our catalogue to sources with a signal-to-noise ratio $\geq 5$, as the catalogue is $\sim$90 per cent complete, and has a false positive rate of less than 1 per cent at this threshold. The bulk of the catalogue sources are previously unknown to the literature, with the majority of unknown sources at sub-mJy levels. Initial science highlights from the catalogue include the detection of 213 radio quiet WISE \hii\ region candidates, previously undetected in radio continuum studies. We show images that compare the SMGPS compact sources to CORNISH ultracompact \hii\ regions, thus highlighting the sensitivity and unprecedented \textit{uv}-coverage of the SMGPS, and the potential synergy of the SMGPS with other surveys.

\end{abstract}




\section{Introduction}
\label{sec:intro}

The SARAO MeerKAT 1.3 GHz Galactic Plane Survey \citep[SMGPS;][]{Goedhart+2024} is one of the most recent deep and high angular resolution radio continuum surveys of the Galactic Plane. SMGPS joins a range of previous and current radio line and continuum surveys at frequencies around a few GHz and with angular resolutions $\sim$1--20\arcsec\ \citep[e.g.][]{White+2005,Hoare+2012,Purcell+2013,Beuther+2016,Green+2017,Brunthaler+2021,Padmanabh+2023,Irabor+2023}, as well as a host of multi-wavelength surveys from the visible to the millimetre wave portions of the electromagnetic spectrum \citep[see][for a comprehensive list of projects]{Molinari+2014,Molinari+2016,Urquhart2024}.

One of the defining data products of many Galactic Plane surveys are point or compact source catalogues, extracted from the survey images or datacubes using automated (or semi-automated) segmentation routines \citep[e.g.][]{Purcell+2013,CSengeri+2014,Molinari+2016,Duarte-Cabral+2021,Medina+2024}. Well constructed catalogues of compact and point sources enable straightforward multi-wavelength analyses and have been crucial in the study of ultracompact \hii\ regions \citep[e.g.][]{Urquhart2013,Kalcheva+2018}, hypercompact \hii\ regions \citep{Yang+2019,Yang+2021,Patel+2023,Patel+2024}, planetary nebulae \citep{Irabor+2018,Fragkou2018}, young stellar objects and radio stars \citep{Umana2012,Lumsden2013,Luque-Escamilla2024}, triggered star formation \citep{thompson+2012, kendrew2012}, the classification of molecular clumps in the Milky Way \citep[e.g.][]{Rigby+2019,Elia2021,Urquhart+2022} and Galactic trends  in  star formation rate and efficiency \citep{djordjevic2019,Rani+2022,zhou2024}. 

SMGPS offers the opportunity to extend these studies into new regions of the depth and angular resolution parameter space; with a sensitivity roughly a factor of 10 greater than the MAGPIS  \citep{White+2005} or THOR surveys \citep{Beuther+2016}, and a resolution a factor of 5 better than the Quadrant IV Molonglo Galactic Plane survey \citep{Green+2012}. In this paper, we present a catalogue of 510\,599 compact radio sources that have been extracted from the SMGPS 1.3 GHz continuum images. In Section \ref{sec:smgps} we describe the pertinent details of the SMGPS observations and data quality. Section \ref{sec:source_extraction} contains a description of our source finding procedure \citep[which uses the \textsc{Aegean} package;][]{Hancock+2012} and the properties of the resulting catalogue (completeness limits, false positive rates, likely artefacts, and source counts). In Section \ref{sec:smgps-simbad}, we speculate on the nature of the SMGPS compact sources and examine their counterparts in the literature. In Section \ref{sec:science}, we present brief science highlights that are enabled by the SMGPS catalogue. Finally, in Section \ref{sec:conclusions}, we summarize our findings and outline our main conclusions.

\section{The SMGPS survey}
\label{sec:smgps}
Comprehensive details of the SMGPS are given in \cite{Goedhart+2024}. Here we give a brief overview of the survey for the convenience of the reader. The SMGPS is an L band survey covering roughly half the Galactic Plane. This survey is spread across two blocks in Galactic longitude of $2^{\circ} \leq l \leq 61^{\circ}$ and $251^{\circ} \leq l \leq 358^{\circ}$, each within a range of approximately $|b|\leq 1.5^{\circ}$ in Galactic latitude. From $l = 300.5^{\circ}$ to $l = 357.5$\degr\ the regions were chosen to account for the warp in the Galactic disc in the same way as the Hi-GAL survey \citep{Molinari+2010}. The SMGPS has an angular resolution of $8''$ and root-mean-square sensitivity of $\sim$10-30 \textmu Jy beam$^{-1}$.

Timing and frequency labelling errors could affect the astrometric accuracy of the data cubes and images. These errors are largely fixed by mosaicing, but are $\sim$0\farcs5 at the edges of the affected mosaics. The Fields affected by these errors are centred at at $l=324.5$ to $l=357.5$ (see Table~\ref{tab:tile_global_rms}). The astrometric accuracy of the observations is estimated at 0\farcs5 for bright sources, depending on their position within the survey \citep[see][for more details]{Goedhart+2024}. The SMGPS flux calibration was compared to the independent THOR survey carried out with the JVLA and found to be accurate within 4 per cent \citep{Goedhart+2024}.

The SMGPS pointing images were restored with the Gaussian fitted to the point spread function (PSF), typically 7.5\arcsec, resulting in similar PSFs and units for the restored and residual components. As part of the mosaic formation, the pointing images were convolved to an effective PSF of 8\arcsec\ and scaled by the change in restoring beam areas. Thus, all the mosaics have an 8\arcsec\ effective PSF.

The SMGPS mosaics (hereafter tiles) are a set of overlapping $\ang{3} \times \ang{3}$ primary beam corrected images. The catalogue data was extracted from images created from the SMGPS data cubes by calculating the zeroth moment. The zeroth moment integrated intensity (hereafter moment 0) images were derived from the data cubes and were calculated on a pixel-by-pixel basis: the summation of the product of the flux density and bandwidth in each fractional bandwidth were weighted by the total bandwidth of all the fractional bandwidth images. Separate catalogues of radio filaments and extended sources are presented in \cite{Williams+2025} and \cite{Bordiu+2025}.

\begin{figure*}
    \includegraphics[width=\textwidth]{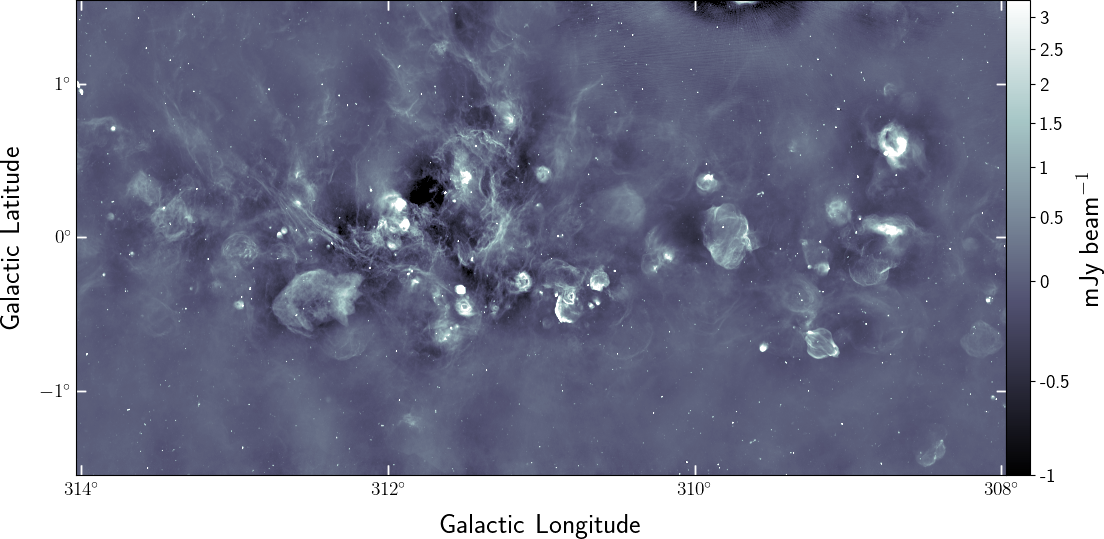}
    \caption{An example image of two SMGPS 1.3 GHz moment 0 tiles, centred at $l=309.5\degr~\&~l=312.5\degr$, that have been stitched together.}
    \label{fig:example_tile_G309.5_G312.5}
\end{figure*}

A sample image of two tiles from the SMGPS is presented in Fig.~\ref{fig:example_tile_G309.5_G312.5}, which shows the moment 0 integrated intensity at 1.3 GHz \citep{Goedhart+2024}. As well as striking emission from extended supernova remnants, \hii\ regions and radio filaments, Fig.~\ref{fig:example_tile_G309.5_G312.5} also reveals a large population of compact radio sources that are smaller than a few times the 8\arcsec\ synthesised beam in diameter.

\section{Method}
\label{sec:method}
\subsection{Source Extraction and cataloguing}
\label{sec:source_extraction}
\begin{figure*}
    \includegraphics[width=\textwidth]{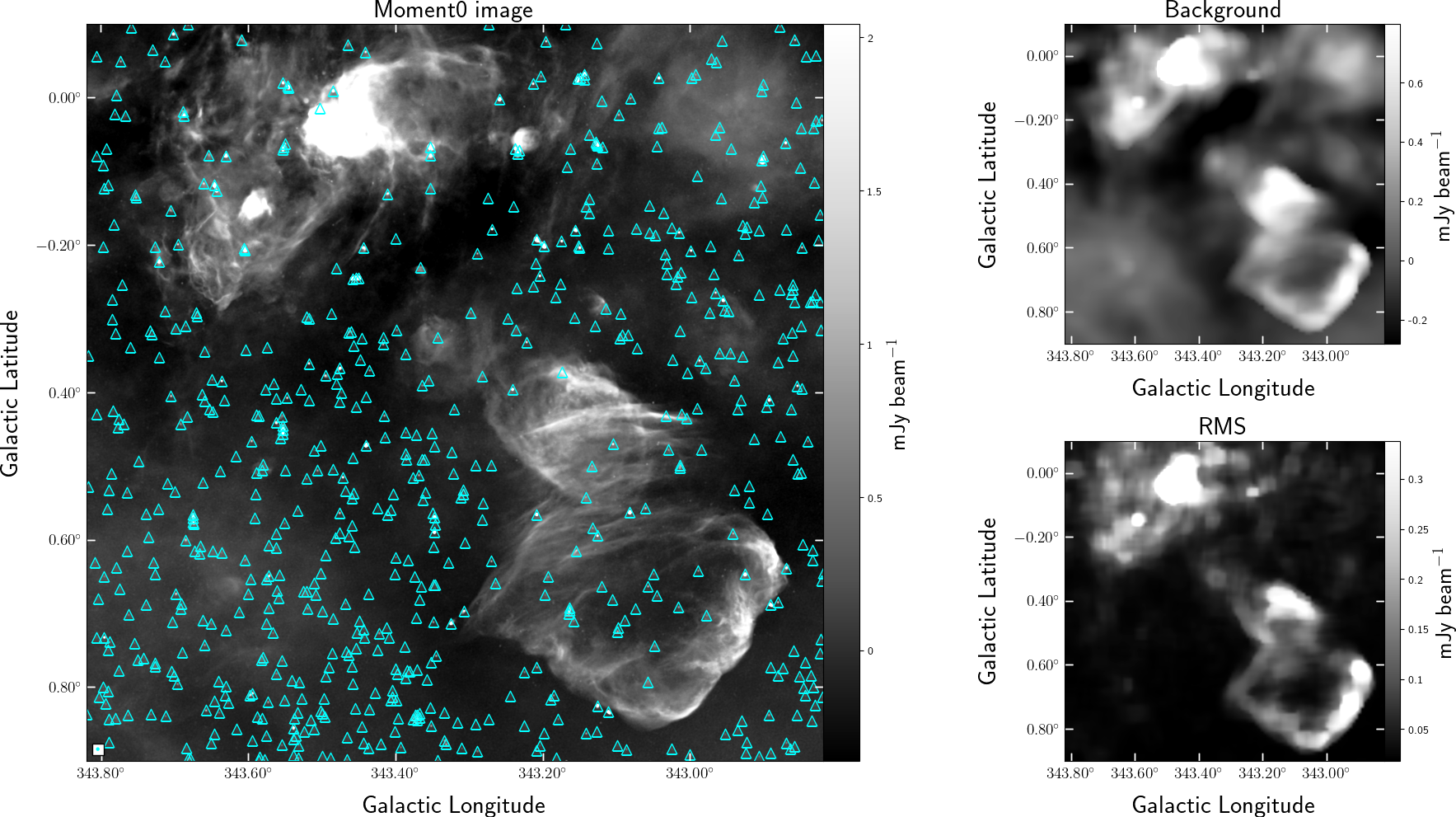}
    \caption{Demonstration of \textsc{BANE}'s background and noise estimation on a square degree of the SMGPS G342.5 tile. The left panel shows the moment 0 image itself, the top-right panel shows the background, and the lower-right panel the rms. Detected sources are marked as cyan triangles on the moment 0 image and the synthesised beam is shown as a filled ellipse in the lower-left corner of the moment 0 image.}
    \label{fig:bane_demonstration}
\end{figure*}

Point (or unresolved) sources are simple to define as an object with an angular size much smaller than the synthesised beam, hence restored in the image as a synthesised beam area object. However, there is no such standard definition for compact sources. Here we define a compact source as one with an area of less than 5 synthesised beams. This is a somewhat arbitrary choice, but one motivated by the decision to identify mainly single component detections. A much larger threshold for size can result in objects being artificially broken up into multiple Gaussian components \citep[e.g. ][]{PurcellHD2007}. Objects with an area greater than 5 synthesised beams are catalogued in the SMGPS extended source catalogue \citep{Bordiu+2025}. In this paper we present the SMGPS compact source catalogue. The background estimation, source identification and characterisation processes were carried out using \textsc{BANE} and \textsc{Aegean} \citep{Hancock+2012, HancockTH2018}. During the background estimation, \textsc{BANE} creates \texttt{background} and \texttt{rms} images (see Fig.~\ref{fig:bane_demonstration}) which are then used to threshold by \textsc{Aegean} during the source extraction.

The background estimation process was carried out using \textsc{BANE} 1.8.0-(2018-09-15). Background estimation is most commonly achieved through a process called thresholding, which separates pixels above a defined threshold (source pixels) from those below the threshold (background pixels). An ideal scenario would present a uniform background to which a single threshold could be applied. However, this is generally not the case. The background and noise properties in the SMGPS fields vary across each tile. \textsc{BANE} uses the \texttt{grid} algorithm, which takes a sliding box-car approach that accounts for this variation across the field. This approach operates on two spatial scales, an inner and outer scale called \texttt{grid} and \texttt{box}, respectively.
The default \texttt{grid} is set to $\sim$4 times the beam, and the default \texttt{box} is 5 times the \texttt{grid}. Both scales are adjustable. Here we have chosen to use the default parameters as they represent a good compromise between computational efficiency and the spatial variation in noise.

The source extraction process was carried out using \textsc{Aegean} 2.0.2-(2018-07-19). \textsc{Aegean} makes use of the \textsc{BANE} results which provide a lower limit for what is considered source emission (See Fig.~\ref{fig:bane_demonstration}).  Pixels identified as source emission are grouped into contiguous groups called islands. For a point source, the island is represented by a single Gaussian component, and for compact sources the island is characterized by a set of Gaussian components. The \textsc{FloodFill} algorithm \citep{Murphy+2007, Hales+2012}, on which \textsc{Aegean} is based, has two thresholds: \texttt{seed} and \texttt{flood}, which are used to identify the brightest pixel in a source and all the adjacent pixels associated with it, respectively. The \texttt{seed} threshold is equal to or greater than the \texttt{flood} threshold as it limits what can be considered a source. We chose to restrict our detections to sources above a signal-to-noise ratio (S/N) of 5, in a similar manner to \cite{Hancock+2012}. However, to ensure sources close to the 5$\sigma$ limit did not go undetected due pixelisation of the images, we set the \texttt{seed} threshold to 4$\sigma$ and manually limited the resulting catalogue to a threshold of 5$\sigma$ in a similar manner to \cite{Hurley-Walker+2017}. The \texttt{flood} threshold limits the growth of the source island, and was kept at its default value of 4$\sigma$. Using these parameters we extracted a total of 763\,902 compact sources with areas less than five 8\arcsec\ beams. After filtering, the catalogue comprises 510\,599 distinct components with a peak-brightness S/N $\geq 5 \sigma$ from 489\,542 islands across 57 survey tiles covering $\sim500$ deg$^2$ of the Galactic Plane.

For each detected component, \textsc{Aegean} performs an iterative Gaussian fitting process from which it produces an extensive listing of every fitted parameter (such as flux density, size and position angle) along with the associated uncertainties, and a report (\texttt{flag}) of the instances where the characterisation failed.

\begin{figure}
	\includegraphics[width=\columnwidth]{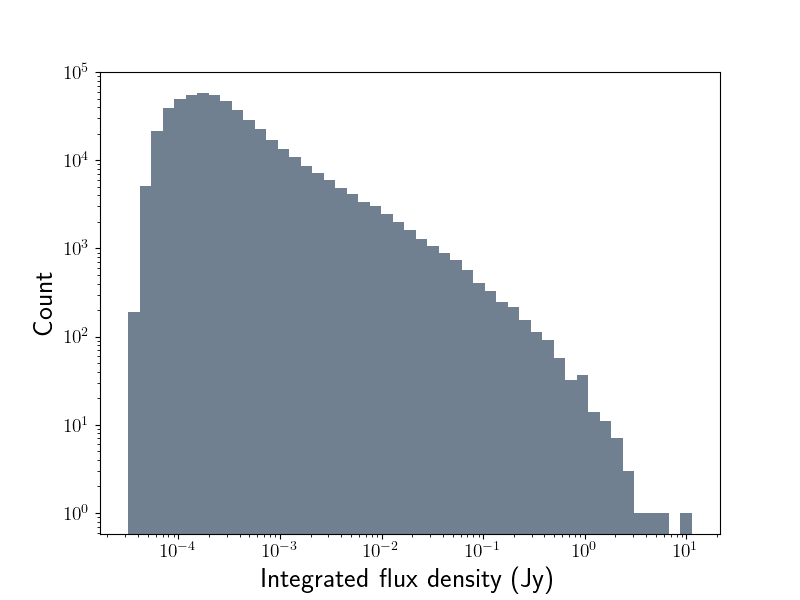}
    \caption{Integrated flux density distribution across the survey. There are 510\,599 sources above 5$\sigma$ with flux densities between 32 \textmu Jy and 11 Jy.}
    \label{fig:integrated_flux_figure}
\end{figure}

\begin{figure}
	\includegraphics[width=\columnwidth]{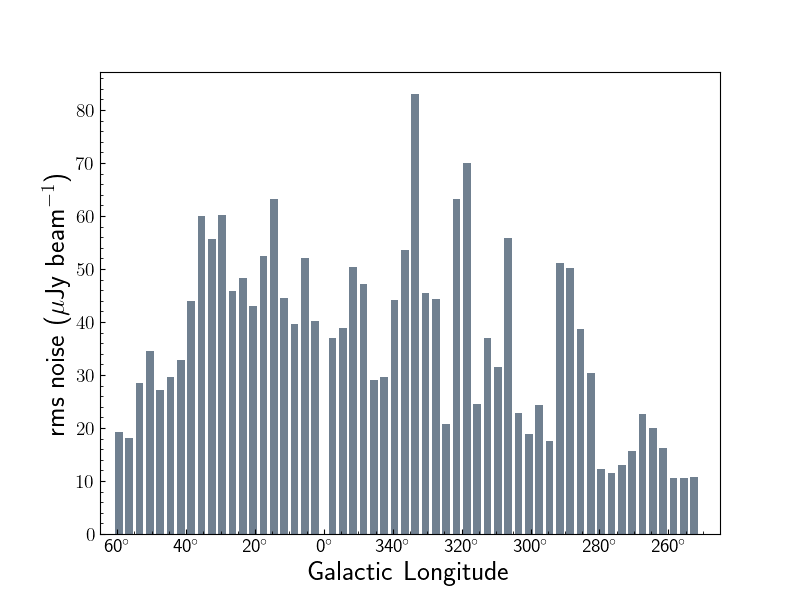}
    \caption{The global rms of each survey tile as a function of Galactic longitude. The global rms is defined as the median value of the pixels in the individual rms images obtained from \textsc{BANE} (see Table~\ref{tab:tile_global_rms}).}
    \label{fig:tile_global_rms}
\end{figure}

The local rms flux density values at the position of the individual sources ranges from 7 \ujy\ to 29 \mjy\ (Fig.~\ref{fig:rms_distribution_figure}) and the integrated flux densities of the extracted sources range from 32 \textmu Jy to 11 Jy (Fig.~\ref{fig:integrated_flux_figure}). The highest rms noise values are coincident with the brightest sources in the catalogue. Sources with a high flux density drive up both local and global rms, the rms noise in the immediate vicinity of the source and the rms noise of the entire image, respectively. As a result the distribution of the values in the \textsc{BANE} images can be highly non-Gaussian, especially in those images containing bright sources, e.g. the G306.5 tile shown in the middle panel of Fig.~\ref{fig:noise_variation}. While the local rms of each detected source is included in the \textsc{Aegean} output, we estimate the global noise of each survey tile from the median value of the pixels in each corresponding \textsc{BANE} \texttt{rms} image (listed in Table~\ref{tab:tile_global_rms} and shown in Fig.~\ref{fig:tile_global_rms}). We refer to this as the ``global rms'' value of each survey tile.

The SMGPS tiles have overlapping regions (see Section~\ref{sec:smgps}) resulting in duplicate sources from adjacent tiles. We filtered the catalogue of duplicates by crossmatching the sources from each tile with those of its overlapping neighbours, and for each duplicate, we removed the source data extracted from the tile with higher global rms.

\subsection{Survey noise}
\begin{figure*}
    \includegraphics[width=\textwidth]{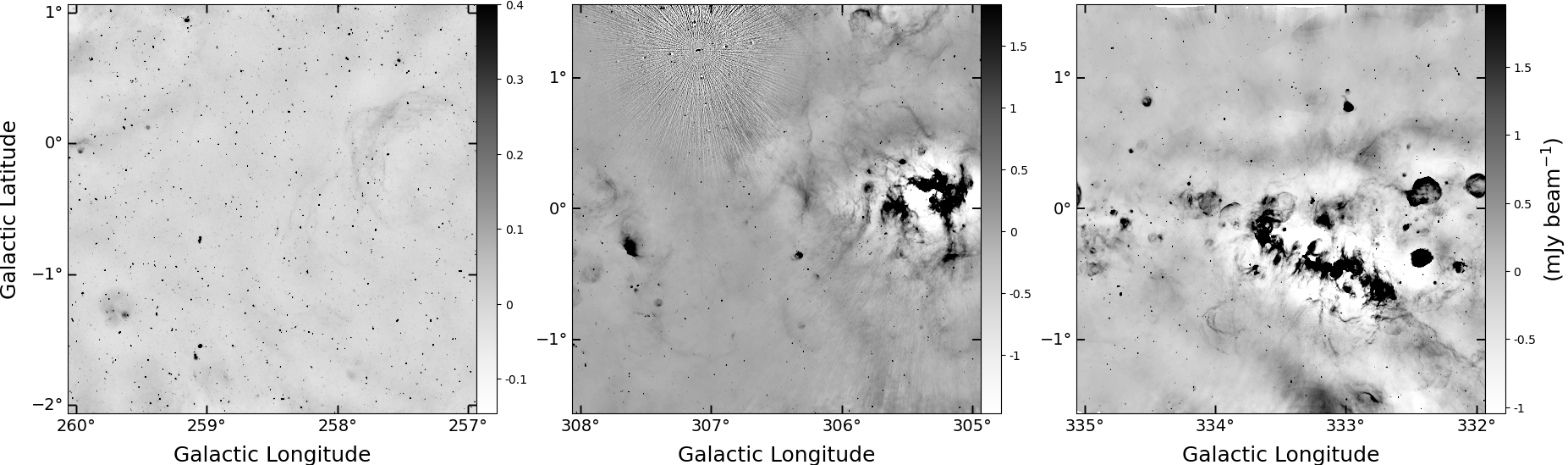}
    \caption{Survey tiles selected to show the local and global variation in noise across the SMGPS. The G258.5 tile (left) is representative of the low-noise portion of the survey, the G306.5 tile (middle) is representative of parts of the survey with huge noise spikes, artefacts, and sidelobes, and the G333.5 tile (right) is representative of parts of the survey with high rms noise due extended defuse emission. The global rms is highest in the G333.5 tile, followed by the G306.5 tile and is lowest in the G258.5 tile: 83, 56 and 11 \textmu Jy beam$^{-1}$, respectively.}
    \label{fig:noise_variation}
\end{figure*}

The difference in background, source density, and brightness of the individual sources between different regions of the survey means there will be spatial variation in noise across the survey. Regions around bright sources, such as quasars, star-forming regions, and \hii\ regions have been found to have higher rms noise values than ones around less bright sources. We show examples in Fig.~\ref{fig:noise_variation} of a well-behaved tile (G258.5: left panel),  a tile hosting a bright quasar (G306.5, middle panel), and a tile dominated by diffuse extended emission (G333.5: right panel). It follows that the G258.5 tile, which does not host bright compact sources, nor extended diffuse emission, is one of the lowest global rms tiles in the survey. Conversely, because of the bright source it hosts, the global rms in the G306.5 tile is five times higher than in G258.5 and the G333.5 tile, which is dominated by diffuse extended emission, has the highest global rms in the survey.

\begin{figure}
	\includegraphics[width=\columnwidth]{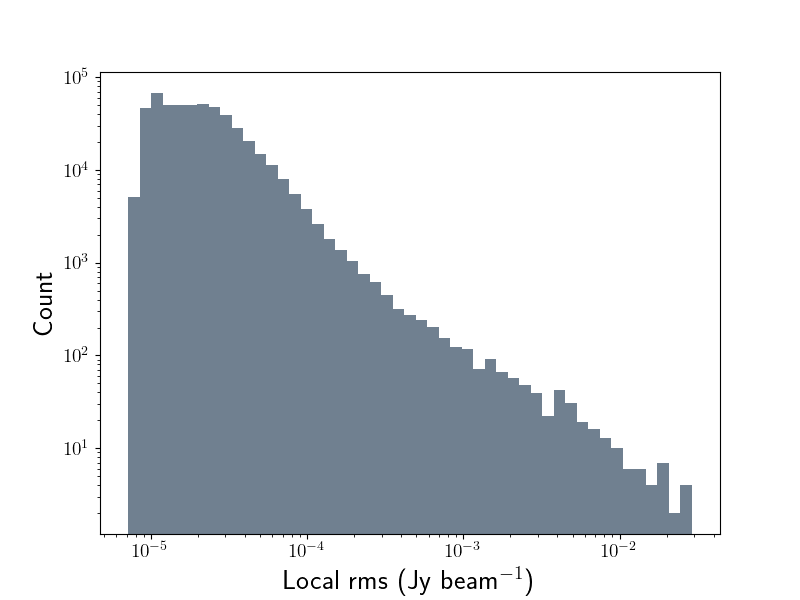}
    \caption{Distribution of the rms noise across the survey, determined through the local rms flux density at the position of the individual SMGPS compact source detections. The local rms values range from 7 \ujy\ to 29 \mjy.}
    \label{fig:rms_distribution_figure}
\end{figure}

Fig.~\ref{fig:rms_distribution_figure} presents the rms noise distribution across the survey. The noise level varies spatially across the survey from 7 \ujy\ to 29 \mjy, with a median value of 19 \ujy. The lower part of the histogram shows a close to Gaussian distribution but the distribution is non-Gaussian at high rms with an extended and asymmetric tail. The highest local rms is seen in the G011.5, G029.5, and G044.5 tiles around sources with flux densities $\geq$ 0.5 Jy, suggesting that both the local and global rms are driven up by source brightness. The highest local rms is seen in the G029.5 tile from the source G030.7675-0.0440 which has a peak brightness of 0.15 Jy beam$^{-1}$ and integrated flux density of 0.55 Jy, indicating that this source is slightly extended, illustrating the varied and highly complex background and rms noise in the SMGPS. 

\begin{table*}
\centering
\caption{SMGPS field rms distribution. Galactic longitude of the field centres are shown in degrees and the global rms (median of \textsc{BANE} rms image) is shown in micro-Janskys per beam. This distribution is shown in Fig.~\ref{fig:tile_global_rms}. Fields centred at $l=324.5$ to $l=357.5$ are affected by timing and frequency errors.}
\label{tab:tile_global_rms}
\begin{tabular}{cc|cc|cc}
\hline
Field Centre                & Global rms        & Field Centre                & Global rms        & Field Centre                & Global rms        \\
Gal. Longitude ($^{\circ}$) & (\textmu Jy beam$^{-1}$) & Gal. Longitude ($^{\circ}$) & (\textmu Jy beam$^{-1}$) & Gal. Longitude ($^{\circ}$) & (\textmu Jy beam$^{-1}$)  \\
\hline
2.5                         & 40                    & 59.5                        & 19                    & 303.5                       & 23                    \\
5.5                         & 52                    & 252.5                       & 11                    & 306.5                       & 56                    \\
8.5                         & 40                    & 255.5                       & 10                    & 309.5                       & 32                    \\
11.5                        & 44                    & 258.5                       & 11                    & 312.5                       & 37                    \\
14.5                        & 63                    & 261.5                       & 16                    & 315.5                       & 25                    \\
17.5                        & 52                    & 264.5                       & 20                    & 318.5                       & 70                    \\
20.5                        & 43                    & 267.5                       & 23                    & 321.5                       & 63                    \\
23.5                        & 48                    & 270.5                       & 16                    & 324.5                       & 21                    \\
26.5                        & 46                    & 273.5                       & 13                    & 327.5                       & 44                    \\
29.5                        & 60                    & 276.5                       & 11                    & 330.5                       & 45                    \\
32.5                        & 56                    & 279.5                       & 12                    & 333.5                       & 83                    \\
35.5                        & 60                    & 282.5                       & 30                    & 336.5                       & 54                    \\
38.5                        & 44                    & 285.5                       & 39                    & 339.5                       & 44                    \\
41.5                        & 33                    & 288.5                       & 50                    & 342.5                       & 30                    \\
44.5                        & 30                    & 291.5                       & 51                    & 345.5                       & 29                    \\
47.5                        & 27                    & 294.5                       & 18                    & 348.5                       & 47                    \\
50.5                        & 35                    & 297.5                       & 24                    & 351.5                       & 50                    \\
53.5                        & 29                    & 300.5                       & 19                    & 354.5                       & 39                    \\
56.5                        & 18                    & 300.5                       & 19                    & 357.5                       & 37              \\
\hline
\end{tabular}
\end{table*}

\subsection{Completeness}
\label{sec:completeness}
\begin{figure*}
    \includegraphics[width=\columnwidth]{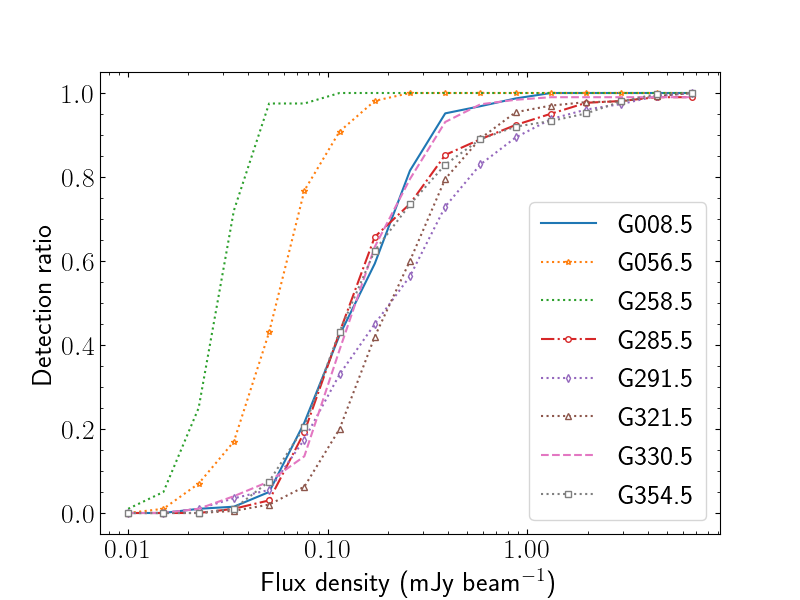}
    \includegraphics[width=\columnwidth]{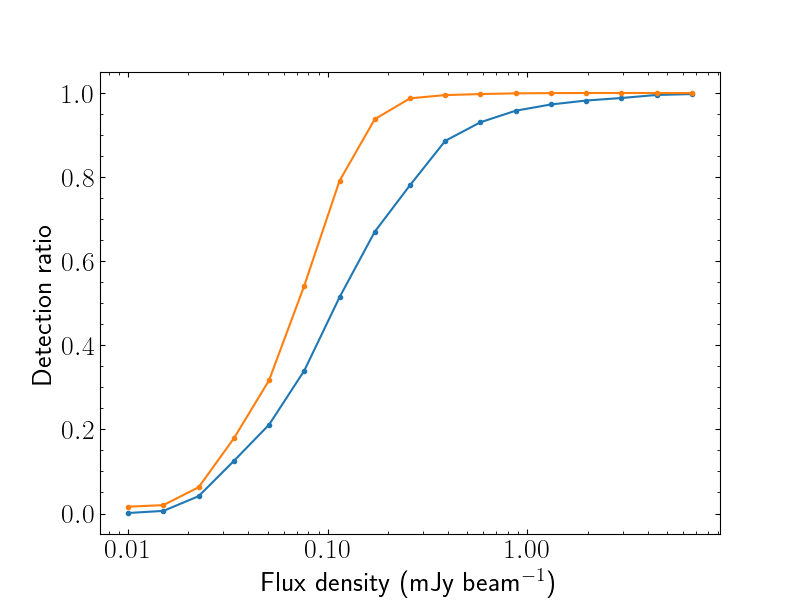}
    \caption{The combined completeness of eight representative tiles as a function of flux density is shown. Completeness of the individual tiles is shown in the left panel. The combined completeness of the respective tiles is shown in the right panel, for the full tiles in blue, and for "well behaved" sub-regions in orange.}
    \label{fig:completeness}
\end{figure*}

The Galactic Plane is a complex environment with varying backgrounds and noise properties (see Fig.~\ref{fig:tile_global_rms}), resulting in an uneven detection of sources across the survey. To examine these effects on the completeness of the resulting compact source catalogue we carried out simulations of source detection for 8 representative tiles. In order to test the source recovery when affected by background and source confusion, the tiles were chosen to have a range of background and source density. For each tile, 1000 realisations were used each of which was injected with 100 randomly generated point sources of the same flux density, for a total of 100\,000 simulated sources in each tile and 800\,000 across the sub-sample. The flux densities, chosen to range from 10 \textmu Jy to 10 mJy to bracket fainter and brighter sources in the catalogue, are different in each realisation but the positions are kept constant. The positions of the simulated sources were randomly generated but not permitted to lie within an arcmin of each other, and were restricted to the central 80 per cent of the respective tiles to avoid edge effects. The simulated sources were catalogued and injected into the moment 0 mosaics using \textsc{AeRes} from the \textsc{Aegean} package. Pre-existing compact source emission was removed from the tiles using \textsc{AeRes}, prior to the artificial-source injection to avoid contamination from pre-existing sources. The \textsc{Aegean} source finder was then run on the artificial-source-injected tiles with the settings used for the original tiles described in section~\ref{sec:source_extraction}. The resulting detections were then compared to the injected sources to estimate the chance of recovering a source as a function of flux density, and thus the completeness of the catalogue. This process was repeated for each representative tile and the combined result is shown in Fig.~\ref{fig:completeness}. The effect of the variation in background and noise across these representative tiles is seen in their respective completeness levels, and is shown in the left panel of Fig.~\ref{fig:completeness}. The combined result is shown in the right panel of Fig.~\ref{fig:completeness} for the full tiles (blue) and for "well-behaved" sub regions (orange) of the respective tiles. We estimate the completeness to be 90 per cent at \s 0.15 mJy beam$^{-1}$ which corresponds to a peak brightness of \s 5$\sigma$. We note that the survey is less complete at this threshold in some parts of the survey as compared to others. We thus do not include the completeness analysis in our decision on the threshold at which to limit our catalogue.

\subsection{Flux boosting}
\label{sec:flux_boosting} 
\begin{figure*}
	\includegraphics[width=\columnwidth]{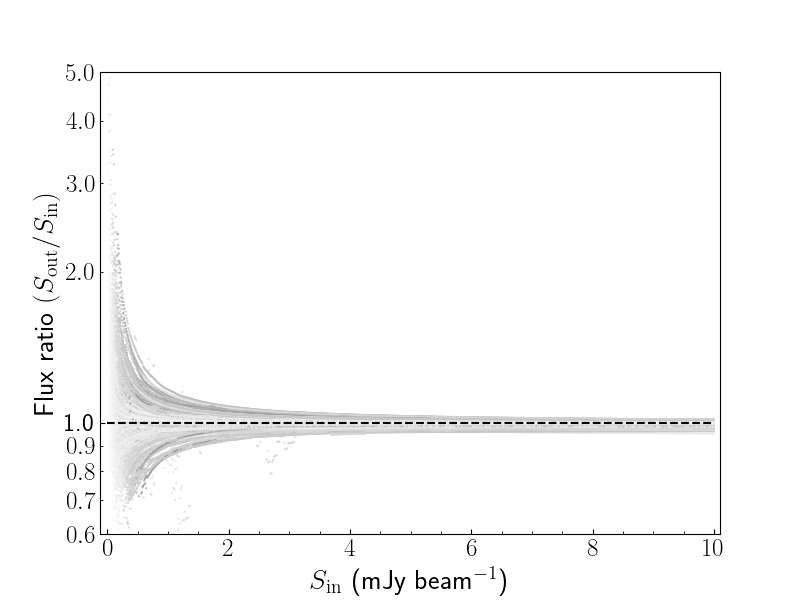}
    \includegraphics[width=\columnwidth]{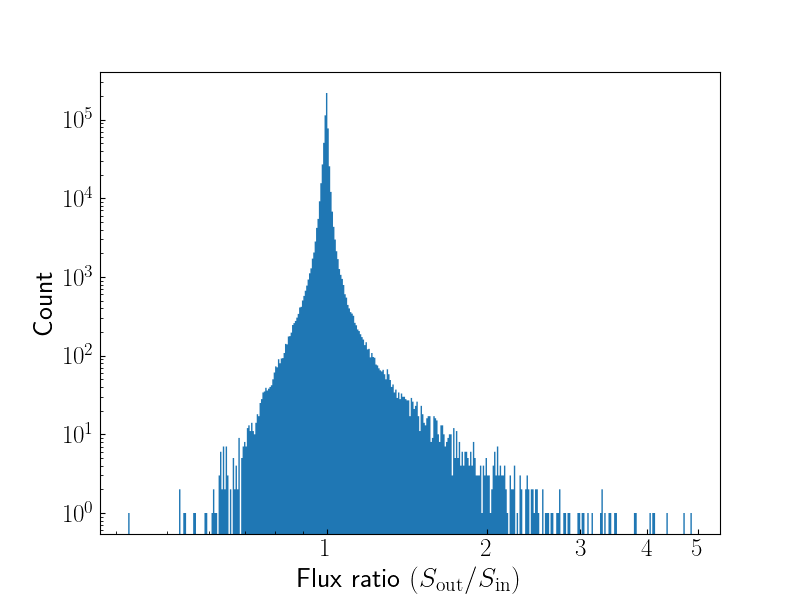}
    \caption{The ratio of recovered flux density to injected flux density for 800\,000 injected sources (100\,000 for each of eight tiles) plotted against the injected flux density. The black dashed line is a line of unity flux density ratio. The plot shows the fainter source flux densities are shifted up when they fall below the noise threshold. In the right panel's histogram the distribution of flux ratios is clustered around 1, showing that there are very few outliers and the majority of the output detections do not deviate from their input value.}
    \label{fig:bias_ratio}
\end{figure*}

The overall accuracy  of the flux scale was assessed in \cite{Goedhart+2024} by comparison against JVLA fluxes from THOR and found to be accurate to within 
$\sim$4 per cent.  In this section the effect of background noise on the detected fluxes is examined by comparison with the simulated catalogues. Due to background and instrumental noise being present in a field, sources with flux densities below the noise threshold that fall on a positive noise spike get shifted up and are thus detected with a higher flux density, an effect referred to as the Eddington bias \citep{Eddington1913}. In Fig.~\ref{fig:bias_ratio} the ratio of detected to input flux density for the simulated source catalogues has been plotted to reveal this ``boosting'' in the detected flux densities. Fig.~\ref{fig:bias_ratio} compiles simulations from the same eight tiles as used in the completeness investigation (section~\ref{sec:completeness}). Each tile was examined separately to investigate potential intra-tile variations. However these were not found and so we present all the simulated sources from the eight tiles in Fig.~\ref{fig:bias_ratio}.

 Fig.~\ref{fig:bias_ratio} clearly shows the flux boosting effect, via an upward trend in the flux ratio at low values of injected flux densities. The effect of noise and/or artefacts on the recovered flux values is generally less than 10 per cent except at low flux densities and/or regions with higher noise. There are a small number ($<$4 per cent of the total number of injected sources) of  outlying points on the graph with anomalously high flux ratios. We investigated each of these anomalous flux ratio sources and in all cases they were either injected on top of diffuse extended emission or into high noise regions like those around bright sources (see the right panel of Fig.~\ref{fig:noise_variation}). 

 As the overall impact of flux boosting is small except at low flux density we have chosen not to apply flux boosting corrections to the compact source catalogue flux densities.

\subsection{False Positive Sources}
\label{sec:false_positives}
\begin{figure}
	\includegraphics[width=\columnwidth]{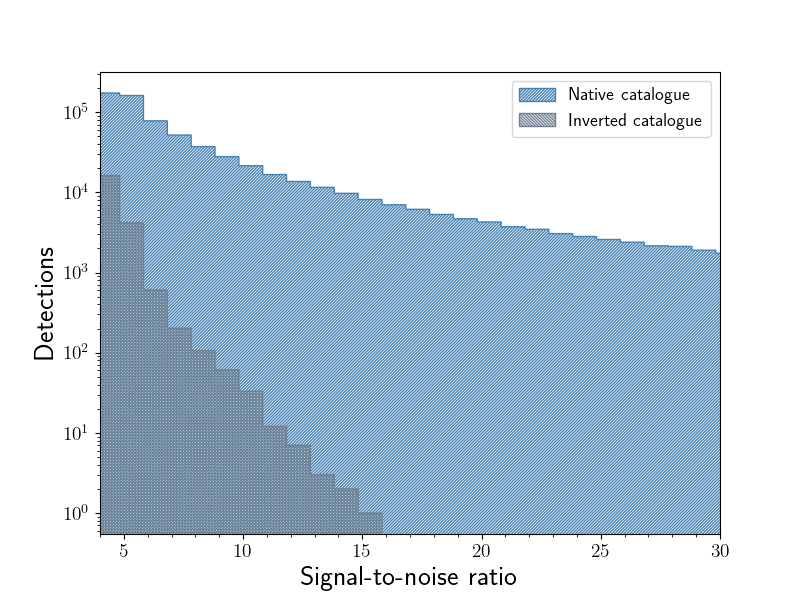}
    \caption{Source counts of the positive and negative detections from the SMGPS moment 0 images as a function of signal-to-noise ratio. The positive sources are shown in the blue histogram, and negative sources in the hatched grey histogram.}
    \label{fig:false_positives}
\end{figure}

\begin{figure}
	\includegraphics[width=\columnwidth]{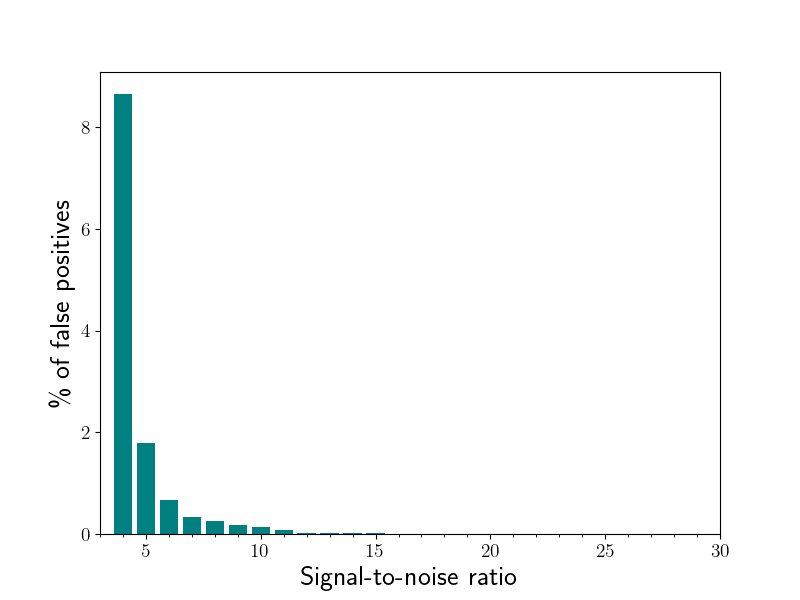}
    \caption{The percentage of false positives is shown as a function of signal-to-noise ratio. The number of false positives detected is under 1 per cent of the full sample. The distribution shows the sample peaking at $\sim 4 \sigma$.}
    \label{fig:false_positives_ratio}
\end{figure}

\begin{figure*}
	\includegraphics[width=\columnwidth]{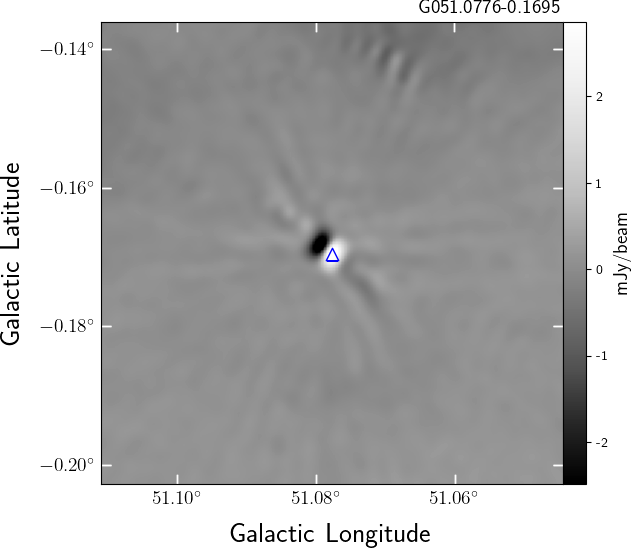}\includegraphics[width=\columnwidth]{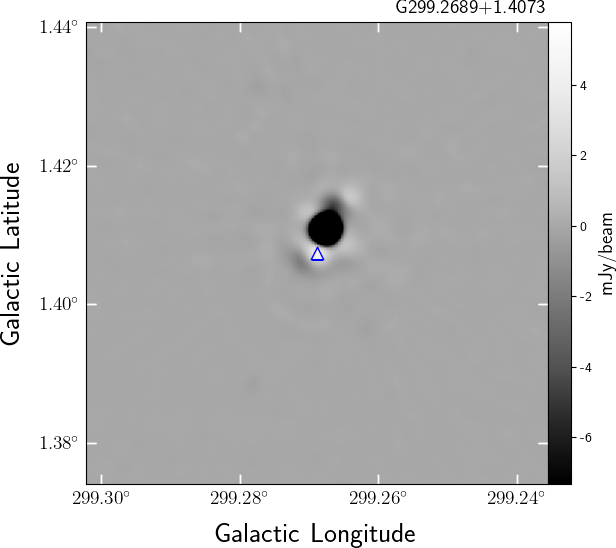}
    \includegraphics[width=\columnwidth]{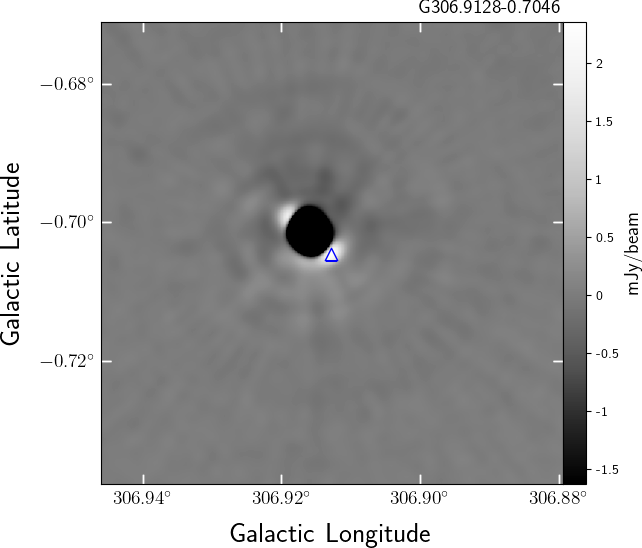}\includegraphics[width=\columnwidth]{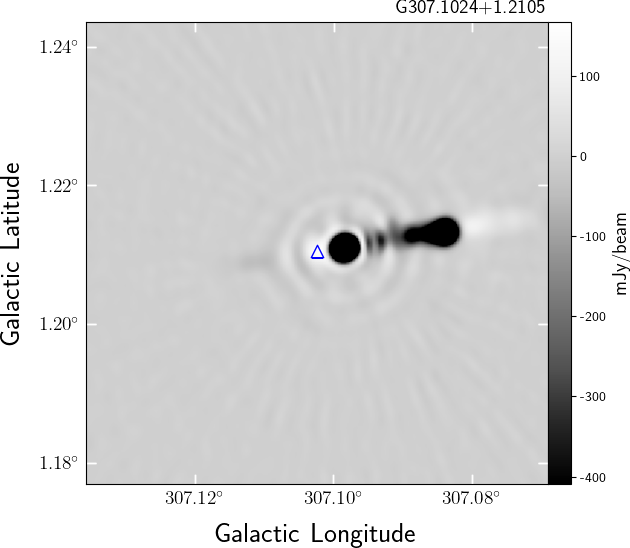}
    \caption{Cutout images of sources from the inverted maps with high S/N. The high-S/N false positives occur as a result of negative bowls around bright sources. The image scale has been stretched to highlight the features being detected as bright high-S/N detections in the inverted maps. The source positions have been marked with blue triangles. Note that the bright sources now appear as negative bowls in these images.}
    \label{fig:negative_bowls}
\end{figure*}

To estimate the rate of false positives, we ran the \textsc{Aegean} source finder on the moment 0 images using the same configuration described in section~\ref{sec:source_extraction}, but with \textsc{Aegean} set to only report detections with negative flux densities. The negative detections represent false positives as we do not expect to detect continuum sources in absorption, and assuming a symmetry about zero for the noise distribution, we expect as many spurious sources as there are negative detections. We invert the negative detections in order to compare them with the standard catalogue and show the comparison in Fig.~\ref{fig:false_positives} as a function of signal-to-noise ratio: the native map detections are shown in the plain blue histogram and the inverted detections are shown in the hatched-grey histogram. A total of 510\,599 sources with areas equivalent to five 8\arcsec\ beams or less were extracted from the moment 0 images with positive peaks above 5$\sigma$. The total number of sources with negative peaks extracted under the same conditions is 3211. Thus, we estimate a false positive rate of 0.63 per cent for sources with S/N $\geq |5\sigma|$. We break this down as a function of signal-to-noise ratio in Fig.~\ref{fig:false_positives}. There are 3211 false positives below $-5\sigma$, 356 of which occur above 7$\sigma$. The most negative detection in S/N was -15.4$\sigma$ with peak intensity 0.59 mJy beam$^{-1}$. Fig.~\ref{fig:false_positives_ratio} shows the result as a percentage. Here we see the highest percentage peaking below 5, with a less than 1 per cent occurrence of false positives at any of the high-S/N values between 7 and 15. The false positives we see at high S/N occur because the noise is highly non-Gaussian as the images are very shallowly cleaned and contain negative bowls around bright sources. Inverting the maps causes these negative bowls to themselves appear as bright sources as shown in Fig.~\ref{fig:negative_bowls}. Therefore, we expect the majority of false positives in our compact source catalogue to occur below an S/N of 5. 

\subsection{The Catalogue}

\begin{figure}
	\includegraphics[width=\columnwidth]{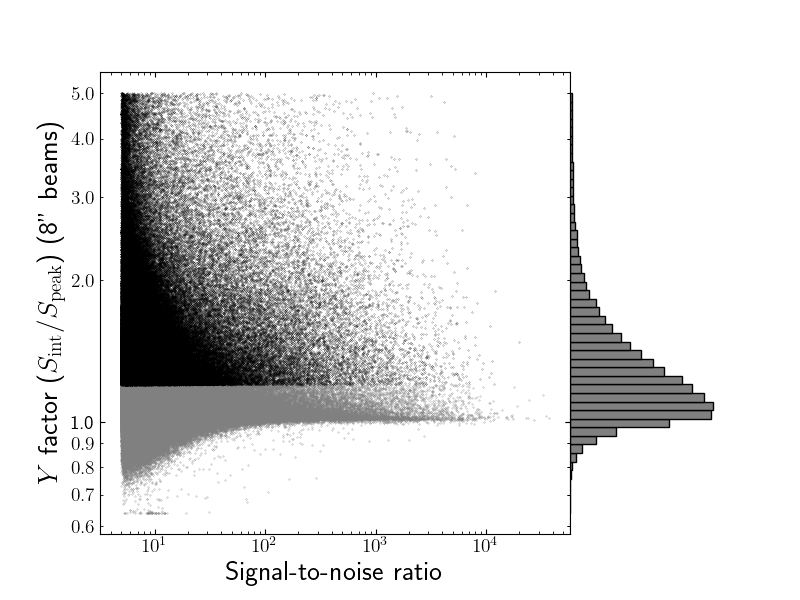}
    \caption{Histogram of SMGPS compact source $Y$-factor (integrated flux divided by peak flux) in units of \ang{;;8} beams.}
    \label{fig:y_factor}
\end{figure}

 The final SMGPS compact source catalogue comprises 510\,599 sources with a peak signal-to-noise ratio $\geq 5$, extracted as described in section~\ref{sec:source_extraction}. An excerpt of the catalogue is shown in Table~\ref{tab:smgps_catalogue_snippet}. The high sensitivity of the SMGPS has produced high dynamic range images, thus some compact sources are parts of larger complexes resolved into multiple components. The catalogue has a total of 489\,542 source islands. Of these, 473\,421 have been detected as single components and 16\,121 have multiple components: 13\,206 two-component islands; 2054 three-component islands, and 861 islands with four or more components. We have examined the morphology of the compact sources detected in the survey by measuring the source elongations through the source finder-estimated Gaussian fits of the half-power diameters along the major and minor axes. We have also measured the 1.3 GHz $Y$-factor (the ratio of integrated to peak flux) for each compact source in the survey, through which we show the distribution of the source sizes in units of 8\arcsec\ beams. We plot histograms of the compact source elongation ($a/b$), and $Y$-factor ($S_{\text{integrated}}/S_{\text{peak}}$) in Figs~\ref{fig:elongation} and \ref{fig:y_factor}, respectively.

The majority of the compact sources show low elongation ratios, with a median elongation of 1.1, indicating the distribution comprises a predominantly spherical or low axis-ratio population. Roughly 50 per cent of the sources have elongation ratios < 1.1, and $\sim$99 per cent of the sources have elongation ratios < 2 (twice as long as they are wide) leaving 3373 ($\sim$0.7 per cent) sources with an elongation ratio > 2. Above a ratio of 2, and particularly above 3, the Gaussian fits appear inconsistent. While some objects with large elongation ratios, such as edge-on Galaxies or double lobed YSOs, might have been characterised correctly, caution should be taken with these objects.

The $Y$-factor provides a measure as to the degree that a particular source is resolved. By definition in radio images the $Y$-factor is 1 for unresolved (i.e.~point) sources. However, in noisy images sources may have a $Y$-factor < 1 ($S_{\text{integrated}}$ < $S_{\text{peak}}$), which is due to the effect of noise on the flux density: a negative noise spike on the integrated flux, a positive noise spike on the peak flux, or sources that sit in a negative bowl, resulting in a suppressed outer wing. Fig.~\ref{fig:y_factor} shows that most of the compact sources in the SMGPS are marginally resolved and peak at a $Y$-factor of $\sim$1.2. To allow for error, we assume point sources have $S_{\textnormal{integrated}} \leq$ 120 per cent of $S_{\text{peak}}$, giving 254\,017 point sources. All sources with 1.2 < $Y$-factor $\leq$ 5 are assumed to be slightly extended and account for the remaining 256\,582 sources in the compact source catalogue: 223\,126 with $Y$-factor < 2, and 33\,456 with $Y$-factor > 2.

We examined the overall sky density around bright sources and found no evidence of source overdensities in their immediate vicinity. Moreover, we found that bright sources are predominantly located in regions of lower source density suggesting that the high noise around bright sources suppresses source detection. There are 165 sources detected in the survey with flux densities > 0.5 Jy. A total of 53\,624 positive sources, and 604 negative sources across the survey are detected within a degree of a 0.5 Jy source. Therefore we estimate a false positive rate of \s1.1 per cent in regions within half a degree of bright (0.5 Jy) sources. We show an example of clustering around bright sources in Fig.~\ref{fig:quasar_sources}. The location of the brightest source is clearly seen at the converging point of the sidelobes. The positive compact sources are shown as blue triangles and the negative sources as orange circles. Here, we only show sources with signal-to-noise ratio $\geq |5\sigma|$ so as to limit the analysis to the threshold used in the final catalogue. We see no correlation between the positive and negative peaks and find that the positive sources are, in general, not coincident with the sidelobes of the central bright source and are therefore likely to be real sources. There are 97 positive, and 9 negative peaks in the selected region, implying a false positive rate of \s10 per cent. However, the case in Fig.~\ref{fig:quasar_sources} is the most extreme in the survey, and has a rate of false positives a factor 10 higher than in other regions hosting bright sources in the survey. Thus, because the estimated rate of false positives in regions with bright sources is only marginally higher than the survey estimate in section~\ref{sec:false_positives}, we choose not to apply filtering methods to reduce the rate of false positives in regions around bright sources \citep[as in e.g.][]{Hurley-Walker+2022, deRuiter+2024}. We have instead flagged all sources in the final catalogue that are within 0.5 degrees of a source with flux density $\geq$0.5 Jy.

\begin{figure}
	\includegraphics[width=\columnwidth]{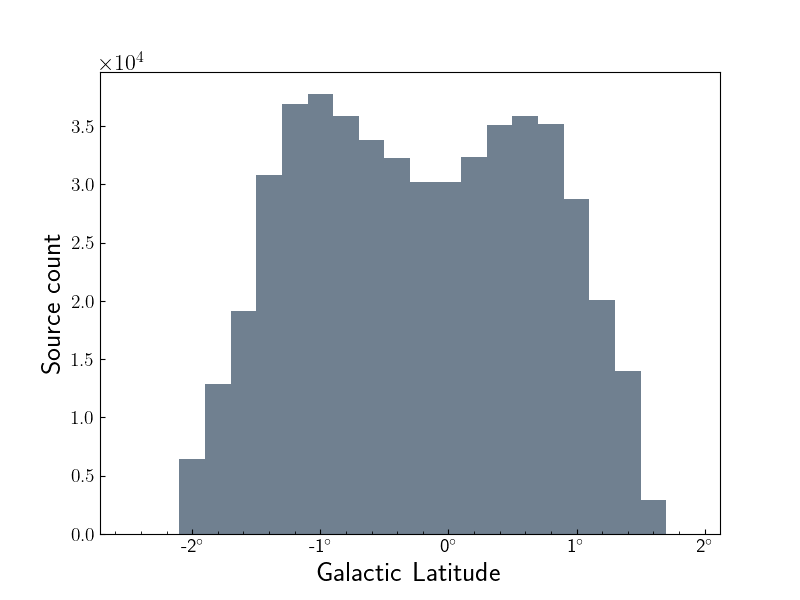}
	\includegraphics[width=\columnwidth]{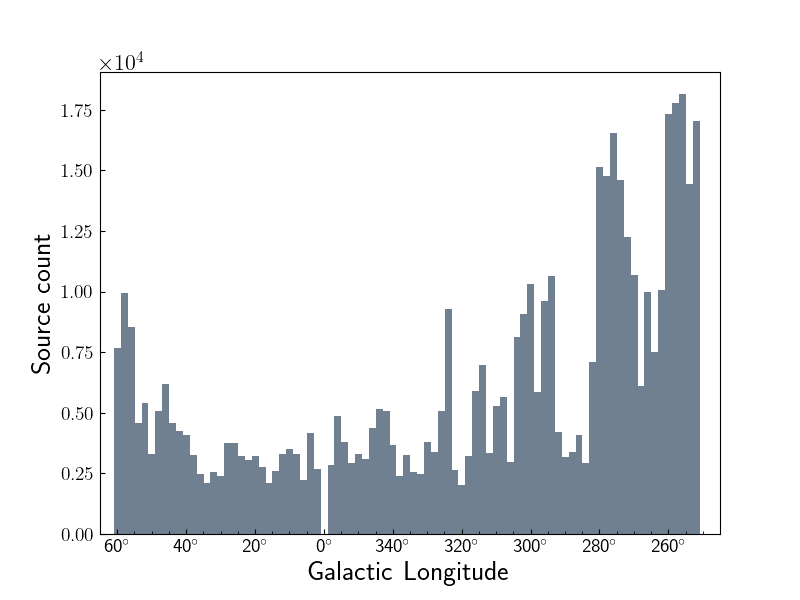}
    \caption{The distribution of source counts across the survey. The upper panel shows the source count as a function of Galactic Latitude and the lower panel shows the distribution as a function of Galactic Longitude. Note that the survey does not cover the Galactic Centre, and only appears to include it because of the bins used.}
    \label{fig:source_count_distribution}
\end{figure}

The distribution of sources as a function of Galactic Longitude and Latitude is shown in Fig.~\ref{fig:source_count_distribution}. The distribution of sources in Galactic Latitude shows a marked decline in source counts toward the mid-Plane. This is likely to be due to extended high surface brightness Galactic foreground emission obscuring background radio galaxies, rather than a true decline. In Galactic Longitude there is an increase in source counts between $l=$ 260\degr\ -- 280\degr, which is due to the Vela region \citep{Rajohnson+2024}. We have used an extragalactic count at the same frequency to estimate the number of chance alignments expected in this population. Scaling from the MIGHTEE \citep{Hale+2023} source counts to the area of the SMGPS, we expect 60 per cent of the sample to be background galaxies.

\textbf{\begin{figure}
	\includegraphics[width=\columnwidth]{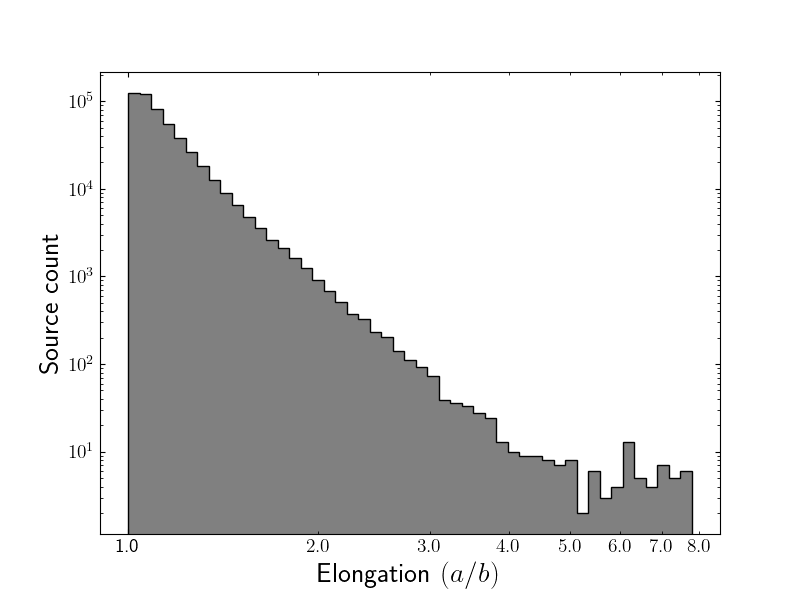}
    \caption{Histogram of SMGPS compact source elongation. Here the ratio of the major to the minor axis is being taken as the source elongation. Circular sources will have a ratio of 1. $\sim$50 per cent of the sources have elongation < 1.1, and $\sim$90 per cent have elongation < 2.}
    \label{fig:elongation}
\end{figure}}

\textbf{\begin{figure}
	\includegraphics[width=\columnwidth]{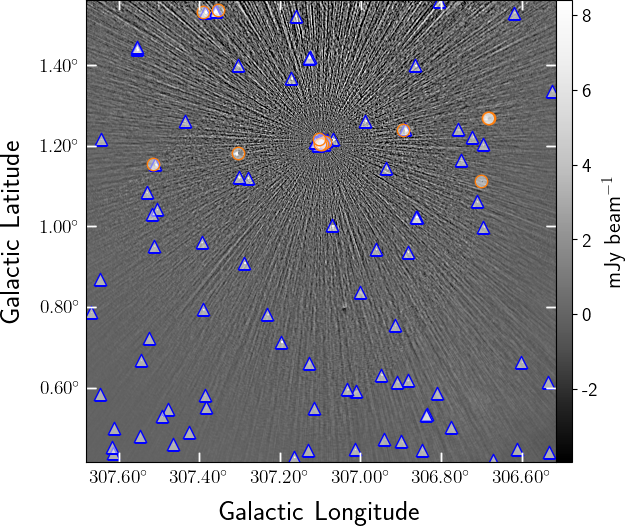}
    \caption{Sources detected within a degree of a bright source. The blue triangles show the positions of 5$\sigma$ and brighter sources. The orange circles show the positions of sources with negative peaks below $-
    5\sigma$. 97 positive peaks and 9 negative peaks are shown in the image.}
    \label{fig:quasar_sources}
\end{figure}}

\begin{table*}
\caption{An excerpt of the SMGPS compact source catalogue. The excerpt shows a subset of the catalogue columns, and the precisions have been truncated for presentation. The full version of the catalogue is available online, and the full range of columns are shown in Table~\ref{tab:smgps_catalogue_format}. Column 1 gives the source identifier; columns 2 and 3, the source positions in Galactic coordinates; columns 4 to 7, the peak intensity, integrated flux density, and their errors; columns 8, 10, and 12 are the major and minor axes of the FWHM, and the PA (with respect to the Galactic coordinates), and columns 7, 9 and 11 are their respective errors.}
\label{tab:smgps_catalogue_snippet}

\centering
\begin{tabular}{llrrlrrrlrlrl}
\hline
Source           & Glon    & \multicolumn{1}{c}{Glat}    & \multicolumn{1}{l}{Peak Intensity}    &       & \multicolumn{1}{c}{$S_{\nu}$} & \multicolumn{1}{l}{} & Major                         &                           & Minor                         &       & PA      &       \\
Galactic name    & (\degr) & \multicolumn{1}{c}{(\degr)} & \multicolumn{1}{l}{(mJy beam$^{-1}$)} & error & \multicolumn{1}{c}{(mJy)}     & error                & \multicolumn{1}{c}{(\arcsec)} & \multicolumn{1}{r}{error} & \multicolumn{1}{c}{(\arcsec)} & error & (\degr) & error \\
\hline
G001.3005+0.1311 & 1.301   & 0.131                       & 16.57                                 & 0.67  & 17.14                         & 0.80                 & 8.28                          & 0.14  & 8.00                          & 0.13  & 7.05    & 0.79  \\
G001.3135+0.3883 & 1.314   & 0.388                       & 12.10                                 & 0.25  & 21.99                         & 0.55                 & 13.48                         & 0.14  & 8.63                          & 0.08  & -0.04   & 0.02  \\
G001.3172+0.3886 & 1.317   & 0.389                       & 2.76                                  & 0.25  & 3.60                          & 0.34                 & 9.61                          & 0.14  & 8.70                          & 0.08  & -86.89  & 0.03  \\
G001.3189+0.2958 & 1.319   & 0.296                       & 1.78                                  & 0.28  & 2.02                          & 0.39                 & 10.01                         & 0.86  & 7.27                          & 0.45  & 20.76   & 0.29  \\
G001.3225+0.2179 & 1.322   & 0.218                       & 8.75                                  & 0.28  & 11.50                         & 0.43                 & 9.36                          & 0.12  & 9.00                          & 0.13  & -82.1   & 0.59  \\
G001.3244-0.1515 & 1.324   & -0.152                      & 8.23                                  & 0.36  & 30.68                         & 1.57                 & 16.71                         & 0.26  & 14.28                         & 0.32  & -74.88  & 0.23  \\
G001.3258+0.3874 & 1.326   & 0.387                       & 2.32                                  & 0.24  & 2.36                          & 0.28                 & 8.47                          & 0.38  & 7.67                          & 0.31  & 10.12   & 0.64  \\
G001.3285-0.0462 & 1.328   & -0.046                      & 5.78                                  & 0.39  & 6.63                          & 0.52                 & 8.64                          & 0.24  & 8.50                          & 0.24  & -85.33  & 2.91  \\
G001.3290+0.1536 & 1.329   & 0.154                       & 1.89                                  & 0.30  & 7.10                          & 1.27                 & 21.18                         & 0.98  & 11.37                         & 0.81  & -67.89  & 0.20  \\
G001.3319+0.0888 & 1.332   & 0.089                       & 15.53                                 & 3.28  & 75.07                         & 17.81                & 18.76                         & 0.98  & 16.49                         & 1.55  & -78.22  & 0.66 \\
\hline
\end{tabular}
\end{table*}

\section{Nature of the SMGPS compact sources}
\label{sec:smgps-simbad}
\begin{figure}
	\includegraphics[width=\columnwidth]{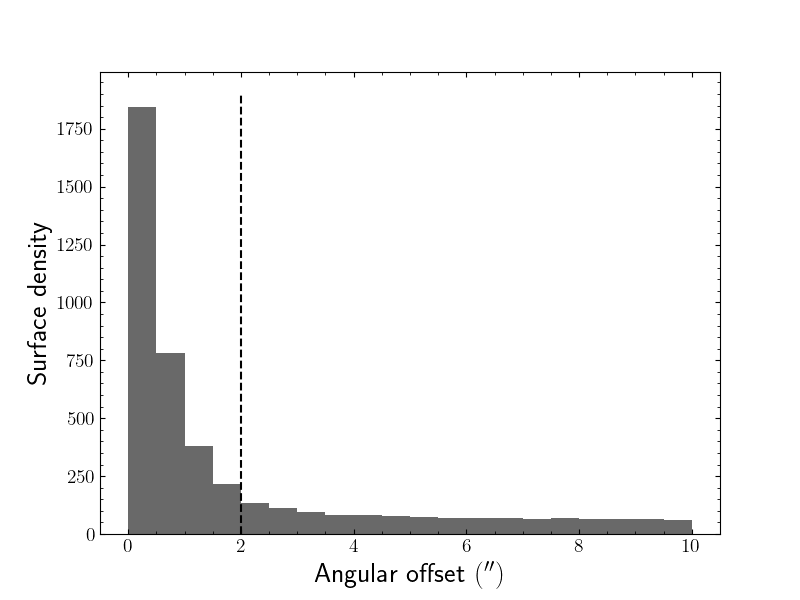}
    \caption{Surface density as a function of angular offset between the SMGPS sources and their respective SIMBAD associations. Matches with an offset of less than 2\arcsec\ are likely to be true associations as these show a high surface density, while those beyond 2\arcsec\ are likely to be chance alignments as the histogram flattens beyond this threshold. The 2\arcsec\ boundary is marked with a vertical dashed line.}
    \label{fig:simbad_offsets}
\end{figure}

In this section we explore the nature of the SMGPS compact sources by associating them with previously classified sources in the literature drawn from the SIMBAD catalogue. The blind matching of SIMBAD sources in such a crowded region as the Galactic Plane against a large radio catalogue is fraught with the potential for chance alignment. We show the probable threshold between true and chance alignments through a plot of surface density as a function of angular offset in Fig.~\ref{fig:simbad_offsets}. For true associations, between the two catalogues, the angular offsets should follow a Gaussian distribution, while a non-Gaussian distribution is indicative of false associations or chance alignments. As the distribution in Fig.~\ref{fig:simbad_offsets} flattens out beyond 2\arcsec\ we expect associations with an offset greater than 2\arcsec\ to likely be false and therefore we only report those associations with an offset $\leq$ 2\arcsec. This is a deliberately cautious choice made in order to report  associations that are as reliable as possible, and as such may miss genuinely true SMGPS-SIMBAD associations, particularly where the astrometric precision of the SIMBAD source is low \citep[for example where single-dish radio detections are reported as in the Parkes-MIT-NRAO survey][]{griffith1991}.

The SMGPS-SIMBAD cross-reference does not serve to conclusively determine the nature of the sources, but to gain a general perspective on what the survey is sensitive to. A complete investigation would require a multi wavelength analysis along the lines of \cite{Urquhart+2018} but this is beyond the scope of this paper. The check returned 92 object types within 2\arcsec, with their distribution shown in Table~\ref{tab:simbad_classifications}. While these object types give an indication of what is detectable in the survey, they represent only $\sim$1 per cent of the full compact source catalogue, thus it would be unwise to speculate the distribution of object types in the full catalogue based entirely on this analysis.

\begin{figure}
	\includegraphics[width=\columnwidth]{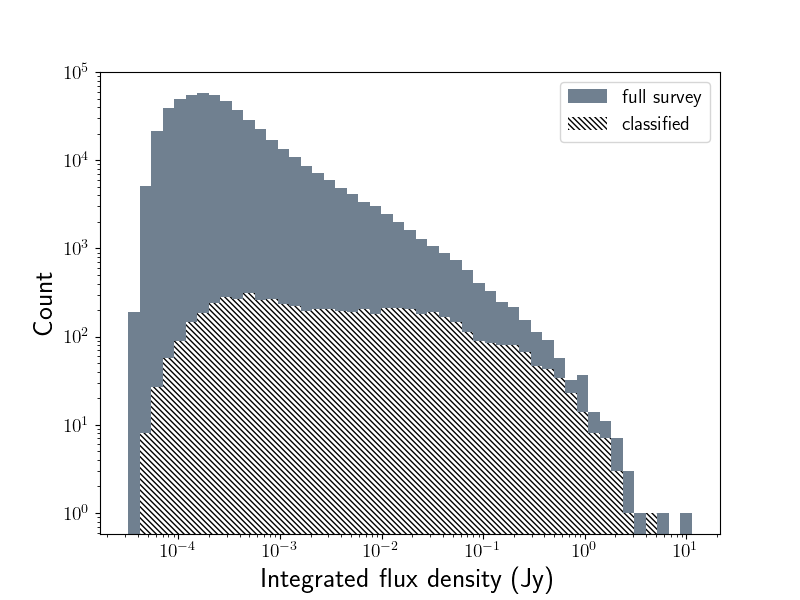}
    \caption{The integrated flux density distributions of the full survey as well as the SIMBAD classified sources are shown in the solid, grey histogram and the white, hatched histograms, respectively.}
    \label{fig:simbad_histogram}
\end{figure}

We show how the SIMBAD-classified flux density distribution compares to the full survey in Fig.~\ref{fig:simbad_histogram}. The figure shows the underlying flux density distribution in the grey histogram overlaid with the flux density distribution of the SIMBAD-classified sources in the white hatched histogram. The flux density distribution of the known objects closely follows the underlying distribution at high flux density but falls off at the lower extreme of the distribution. Approximately 62 per cent of the sources with flux densities above 100 mJy are known while the bulk of the unidentified population is at sub-mJy levels, underlining the importance of such a deep survey for discovering new objects. These objects are unlikely to have been detected prior to SMGPS due to the sensitivity of previous surveys being around a mJy \citep[e.g.][]{Helfand+2006, Hoare+2012, DellerM2014, Beuther+2016}.

\begin{table*}
\caption[Distribution of SIMBAD object types associated with the SMGPS sources.]{Distribution of object types associated with the SMGPS sources. The associations were determined through cross matching the SMGPS compact source catalogue with the SIMBAD database using a 2\arcsec\ matching radius.}
\label{tab:simbad_classifications}

\centering
\begin{tabular}{cccccccc}
\hline
SIMBAD Otype                   & Count & SIMBAD Otype                 & Count & SIMBAD Otype               & Count & SIMBAD Otype                   & Count \\
\hline
Radio                  & 1267  & NearIR               & 14     & gamma                & 5      & FarIR                  & 2      \\
YSO\_Candidate          & 1173  & BlueSG               & 13     & Cluster*\_Candidate   & 5      & Blend                  & 1      \\
smmRad                 & 523   & AGB*                 & 12     & WhiteDwarf           & 4      & Supergiant             & 1      \\
Pulsar                 & 402   & RRLyrae              & 11     & BYDraV*              & 4      & ClG                    & 1      \\
LongPeriodV*\_Candidate & 393   & SB*                  & 10     & RSCVnV*              & 4      & XrayBin                & 1      \\
Star                   & 360   & Outflow\_Candidate    & 10     & AGN\_Candidate        & 4      & Association            & 1      \\
PlanetaryNeb           & 210   & HighMassXBin         & 8      & ISM                  & 4      & S*                     & 1      \\
YSO                    & 174   & MolCld               & 8      & Eruptive*            & 3      & RedSG\_Candidate        & 1      \\
HIIReg                 & 165   & PulsV*               & 8      & gammaDorV*           & 3      & StarFormingReg         & 1      \\
PlanetaryNeb\_Candidate & 158   & RedSG                & 8      & ClassicalCep         & 3      & delSctV*               & 1      \\
OH/IR*                 & 158   & post-AGB*\_Candidate  & 8      & BLLac                & 3      & YellowSG               & 1      \\
EclBin                 & 156   & mmRad                & 7      & alf2CVnV*            & 3      & C*                     & 1      \\
MidIR                  & 91    & RadioG               & 7      & Variable*            & 3      & OpenCluster            & 1      \\
Infrared               & 79    & DarkNeb              & 7      & C*\_Candidate         & 3      & RefNeb                 & 1      \\
denseCore              & 62    & cmRad                & 7      & SNRemnant            & 3      & Type2Cep               & 1      \\
LongPeriodV*           & 47    & **                   & 7      & Galaxy\_Candidate     & 3      & BlueSG\_Candidate       & 1      \\
Bubble                 & 40    & OrionV*              & 7      & Outflow              & 3      & QSO                    & 1      \\
Galaxy                 & 39    & RotV*                & 6      & Blazar\_Candidate     & 3      & HerbigHaroObj          & 1      \\
AGB*\_Candidate         & 36    & Nova                 & 6      & Symbiotic*\_Candidate & 2      & Blazar                 & 1      \\
EmObj                  & 36    & post-AGB*            & 6      & Seyfert2             & 2      & HighMassXBin\_Candidate & 1      \\
X                      & 32    & Symbiotic*           & 6      & Unknown              & 2      & Seyfert1               & 1      \\
WolfRayet*             & 27    & RGB*                 & 6      & Cloud                & 2      & GlobCluster            & 1      \\
Maser                  & 25    & WhiteDwarf\_Candidate & 6      & EllipVar             & 2      &                        &        \\
Mira                   & 20    & HighPM*              & 5      & PartofCloud          & 2      &                        &        \\
EmLine*                & 14    & Be*                  & 5      & Cluster*             & 2      &                        &        \\

\hline
\end{tabular}
\end{table*}

\begin{figure*}
    \includegraphics[scale=0.33]{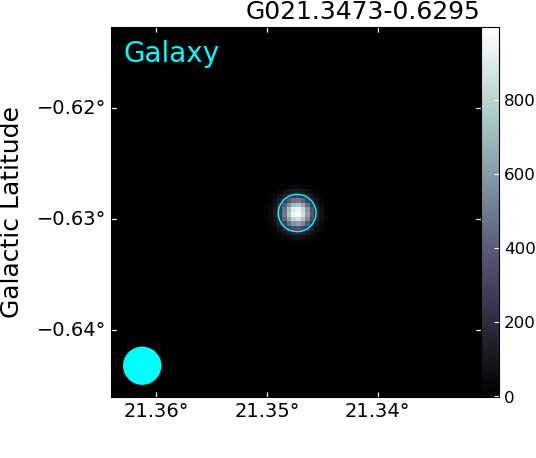}\includegraphics[scale=0.33]{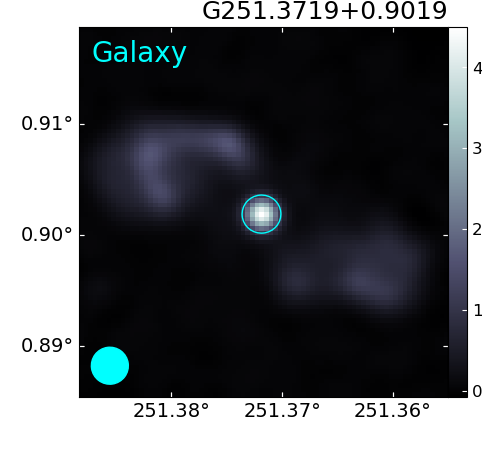}\includegraphics[scale=0.33]{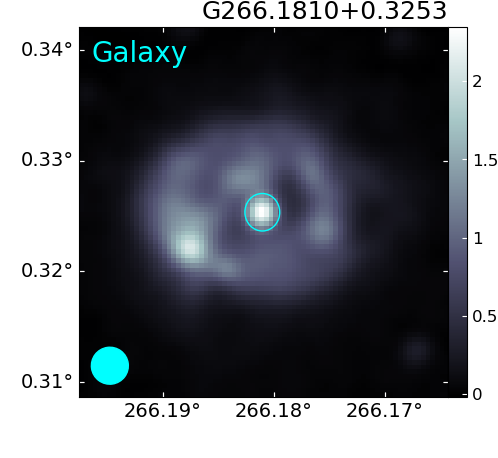}\includegraphics[scale=0.33]{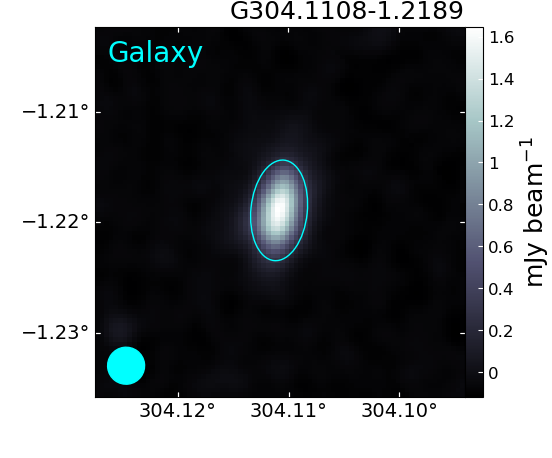}
    \includegraphics[scale=0.33]{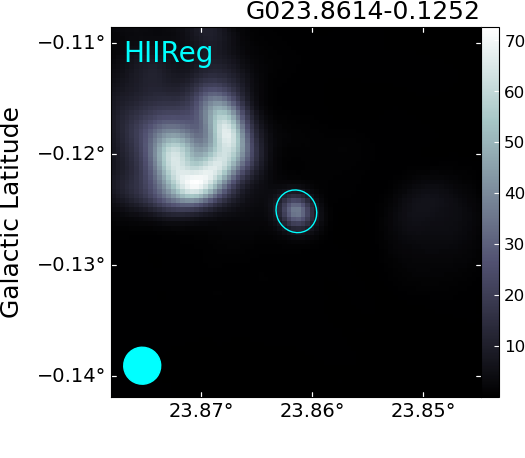} \includegraphics[scale=0.33]{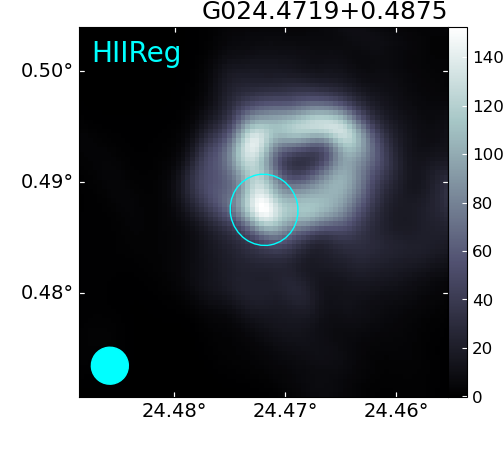}\includegraphics[scale=0.33]{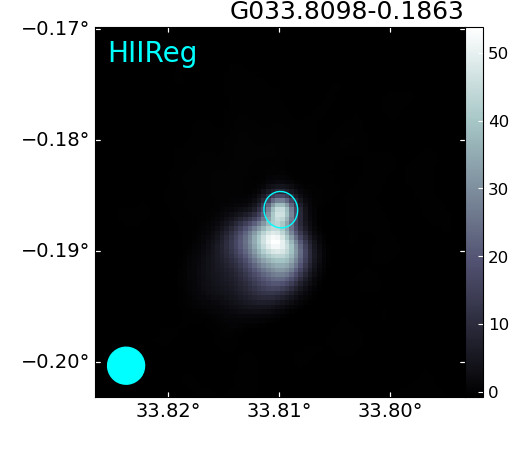} \includegraphics[scale=0.33]{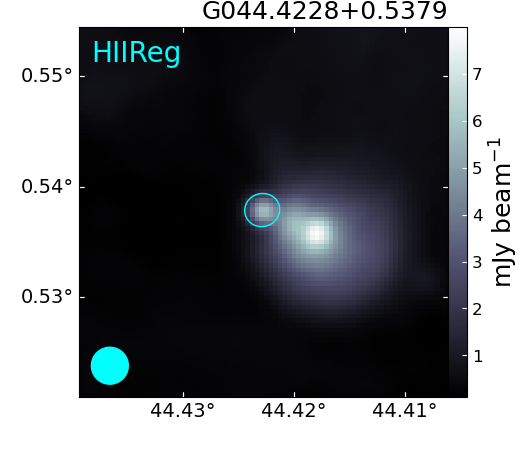}
    \includegraphics[scale=0.33]{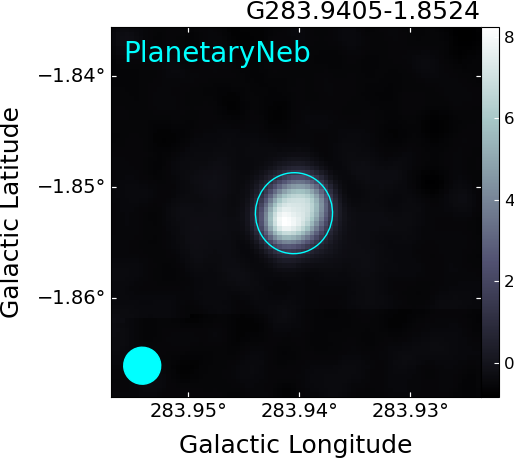} \includegraphics[scale=0.33]{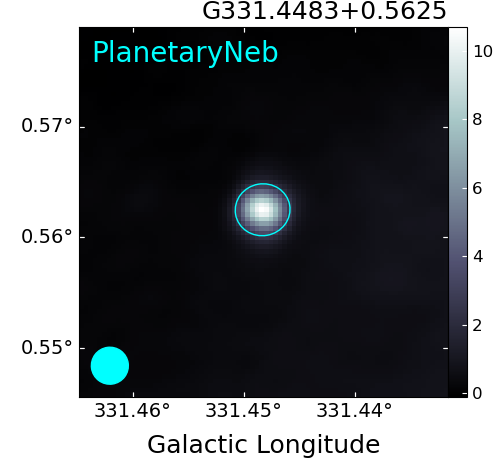} \includegraphics[scale=0.33]{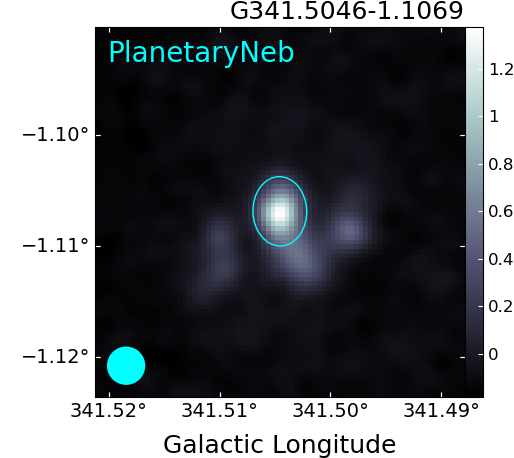} \includegraphics[scale=0.33]{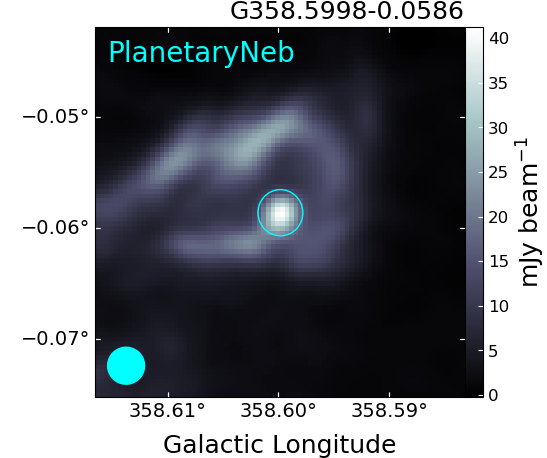}
    \caption{Example images of galaxies, \hii\ regions, and planetary nebulae in the SMGPS classified through the SIMBAD database. The filled circle in the lower left corner of the images represents the SMGPS \ang{;;8} beam, and the empty ellipses encompass the emission being detected as a compact source in the respective images. A summary of the source parameters is shown in Table~\ref{tab:example_images}.}
    \label{fig:simbad_sources}
\end{figure*}

There are 5985 SIMBAD-classified sources represented in this distribution of 92 object types. We have selected a sub-sample from the galaxy, \hii\ region, and planetary nebulae (PNe) classes to present as example images, shown in Fig.~\ref{fig:simbad_sources}. These classes are adequately representative of the differences in morphology between object types throughout the survey as well as the differences between objects with the same classification. A summary of the parameters of these objects is shown in Table~\ref{tab:example_images}.

We see 39 sources matching a known galaxy, 4 of these are presented in the first row of Fig.~\ref{fig:simbad_sources}. G021.3473-0.6295 is a beam-sized detection and is representative of a bright point source. G251.3719+0.9019 and G266.1810+0.3253 have more complex and resolved structure. However, in these, our compact source detection is a core embedded within the complex structure. While in G304.1108-1.2189 our detection is marginally resolved and extends to a few beams in size.

Examples of the detections in the \hii\ region class are shown in row 2 of Fig.~\ref{fig:simbad_sources}. These have been chosen to highlight the varied morphologies of \hii\ regions as they evolve. The images show the source being detected as a compact core associated with diffuse, extended emission which we expect for \hii\ regions that have recently formed due massive stars ionising the gas in their ambient medium.

In a similar manner to the galaxies and \hii\ regions, the PNe show a range of morphologies, from point sources to marginally resolved compact sources. Example images of objects in the PNe class are shown in the third row of Fig.~\ref{fig:simbad_sources}. We see both isolated compact sources and sources associated with diffuse emission. Evidently, the sensitivity and depth of the SMGPS makes it possible to study these objects over a wide range of morphologies.

\begin{table*}
\caption{Summary of the parameters of the objects in the example images of sources from the SMGPS catalogue with SIMBAD classified associations shown in Fig.~\ref{fig:simbad_sources}. A subset of the star, galaxy, \hii\ region, and Planetary Nebulae classes has been chosen to be representative of the varying morphologies.}
\label{tab:example_images}
\centering
\begin{tabular}{llllll}
\hline
Source           & \multicolumn{2}{c}{Observed Flux Density} & Nearest SIMBAD ID             & Angular Dist. & Otype  \\
Galactic name    & Peak (mJy beam$^{-1}$)              & Integrated (mJy)           &                               &($''$)               &        \\
\hline
G021.3473-0.6295 & 1007.18    & 1038.17   & ICRF J183220.8-103511     & 0.75                  & Galaxy       \\
G251.3719+0.9019 & 4.31       & 4.65      & 2MASX J08143768-3305480   & 0.41                  & Galaxy       \\
G266.1810+0.3253 & 1.66       & 1.59      & ZOA J085828.676-451630.99 & 0.47                  & Galaxy       \\
G304.1108-1.2189 & 1.69       & 7.04      & ZOA J130213.424-640356.57 & 1.3                   & Galaxy       \\
G023.8614-0.1252 & 36.25      & 45.97     & CORNISH G023.8618-00.1250 & 0.42                  & HIIReg       \\
G024.4719+0.4875 & 113.59     & 401.23    & CORNISH G024.4721+00.4877 & 0.95                  & HIIReg       \\
G033.8098-0.1863 & 32.7       & 29.37     & IRAS 18511+0038           & 0.88                  & HIIReg       \\
G044.4228+0.5379 & 3.85       & 3.27      & CORNISH G044.4228+00.5377 & 0.57                  & HIIReg       \\
G283.9405-1.8524 & 9.52       & 43.48     & PN Hf    4                & 1.29                  & PlanetaryNeb \\
G331.4483+0.5625 & 10.12      & 21.07     & PN Pe  1-4                & 0.66                  & PlanetaryNeb \\
G341.5046-1.1069 & 1.37       & 3.73      & PN G341.5-01.1            & 0.72                  & PlanetaryNeb \\
G358.5998-0.0586 & 33.81      & 51.25     & JaSt 37                   & 0.64                  & PlanetaryNeb \\
\hline
\end{tabular}
\end{table*}

\section{Science highlights}
\label{sec:science}

\subsection{Radio quiet compact \hii\ region candidates from the WISE catalogue}
The WISE catalogue of \hii\ regions \citep{Anderson+2014} comprises 8399 sources across five categories: known sources (K) are those with measured radio recombination lines or H$\alpha$ emission; groups (G) are \hii\ region candidates located within the same complex as a known \hii\ region; candidate sources (C) are those associated with radio continuum emission but have neither radio recombination line nor H$\alpha$ observations; sources for which there is no detected radio continuum emission in previous surveys are radio quiet (Q); and sources not in any of the aforementioned groups are unclassified.

\begin{figure}
    \includegraphics[width=\columnwidth]{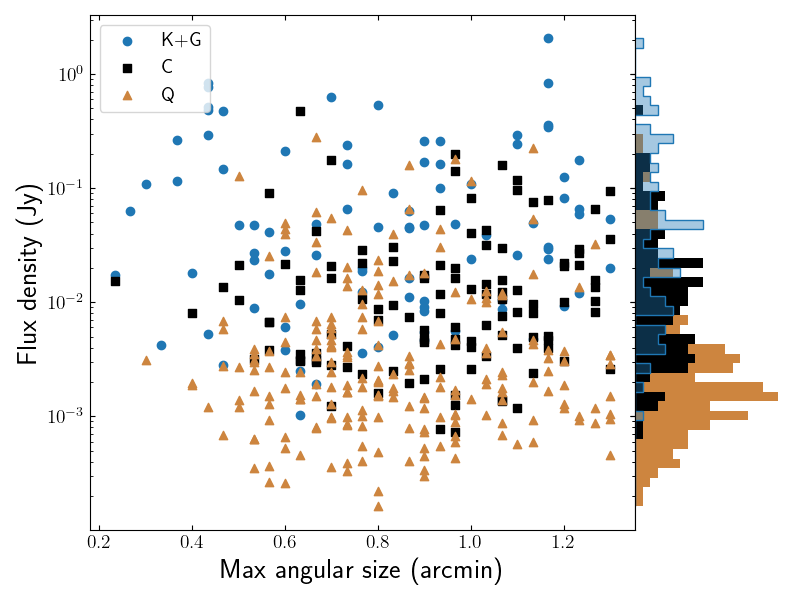}
    \caption{SMGPS flux density of compact \hii\ regions in the WISE \hii\ region catalogue as a function of their WISE angular size. Four categories are shown following the definitions in the WISE \hii\ region catalogue: 'K', 'G', 'C', 'Q', for Known, Groups, Candidates, and Radio quiet, respectively.}
    \label{fig:wise_hii_regions}
\end{figure}

\cite{Bordiu+2025} investigated the association of the radio quiet population with radio continuum emission from the SMGPS extended source catalogue (source areas > five 8\arcsec\ beams), and found extended emission associated with 759 radio quiet sources. Similarly, here, we investigate the association of compact radio sources with the compact (radius $<40\arcsec$) radio quiet population of the WISE \hii\ region catalogue. There are 2255 WISE \hii\ region candidates with radii less than 40\arcsec, of these 2001 are within the SMGPS area. We detected 418 compact radio sources associated with a source in the WISE catalogue of \hii\ regions. Of these, 213 are radio quiet sources, 115 are candidates, 48 are groups, and 42 are known sources. We show the flux densities of the associated sources as a function of their angular size in Fig.~\ref{fig:wise_hii_regions}. We find that the radio quiet detections are predominantly fainter sources which follows the trends shown in \cite{Umana+2021} and \cite{Bordiu+2025}, so they are likely to have been classified by \cite{Anderson+2014} as radio quiet because they are fainter and would have gone unseen in previous radio studies.

While these previously radio quiet objects are now associated with radio emission, our speculation on their nature remains uncertain. The compact population of the WISE \hii\ region candidates could potentially be contaminated with planetary nebulae and wind-blown bubbles. A study, using higher resolution infra-red data, to investigate SEDs and colours is necessary to be sure of their nature, as has been done previously for RMS \citep{Lumsden2013} and CORNISH \citep{Purcell+2013, Irabor+2023} samples. The multi-wavelength analysis required to ascertain the true nature of these \hii\ region candidates, which is beyond the scope of this paper, will be carried out in a future study. 

\subsection{CORNISH ultracompact \hii\ regions in the SMGPS}
\label{sec:cornish_hiis}
The Coordinated Radio and Infrared Survey for High-Mass Star Formation (CORNISH: \citealp{Hoare+2012}; \citealp{Purcell+2013}), is an arcsecond resolution radio continuum survey of the Inner Galactic plane observed by the Very Large Array (VLA) in B and BnA configuration at 5 GHz. CORNISH covers the same region as, and has a similar resolution ($\ang[angle-symbol-over-decimal]{;;1.5}$) to the northern $Spitzer$ GLIMPSE I region; $10^{\circ}<l<65^{\circ}$ and $|b|<1^{\circ}$. Driven by the goal of studying the formation of massive stars, this is a blind survey aimed at thermal sources such as UC \hii\ regions. The survey has detected 3062 sources above a 7$\sigma$ detection limit, with greater than 90 per cent completeness at a flux density of 3.9 mJy.

\begin{figure}
    \includegraphics[width=\columnwidth]{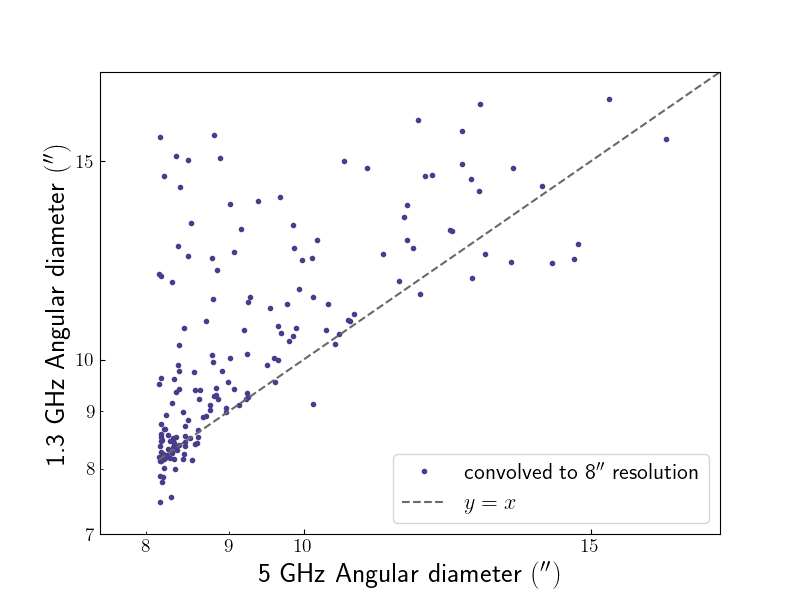}
    \caption{Size-size comparison between CORNISH 5 GHz UC \hii\ regions and their 1.3 GHz SMGPS counterparts. The CORNISH sources have been convolved to an 8\arcsec\ beam to enable a like-for-like comparison with the SMGPS sources. The line of equality is marked with a dashed line.}
    \label{fig:diam_comparisons}
\end{figure}

\cite{Kalcheva+2018} present a catalogue of 239 UC \hii\ regions found in the CORNISH survey, 176 of these are seen in the SMGPS compact source catalogue. We have used these \hii\ regions to draw a comparison between UC \hii\ regions in the CORNISH survey and compact sources in the SMGPS. The synthesized beam in the 5 GHz CORNISH observations is 1\farcs5, and the SMGPS, observed at 1.3 GHz, has a resolution of 8\arcsec. MeerKAT has significantly better \textit{uv}-coverage than the snapshot CORNISH images taken by the VLA, resulting in much better sensitivity on large angular scales \citep[27\arcmin\ vs 14\arcsec][]{Bordiu+2025,Purcell+2013}. Fig.~\ref{fig:diam_comparisons} shows a comparison between the angular diameters of sources within the two surveys, with the CORNISH sources convolved to the same 8\arcsec\ resolution as the SMGPS ones. The SMGPS sources tend to be larger than the CORNISH ones.

\begin{figure*}
    \includegraphics[width=\columnwidth]{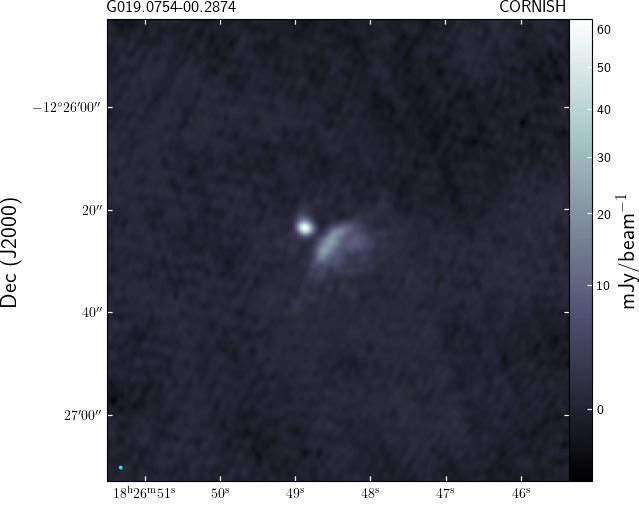}\includegraphics[width=\columnwidth]{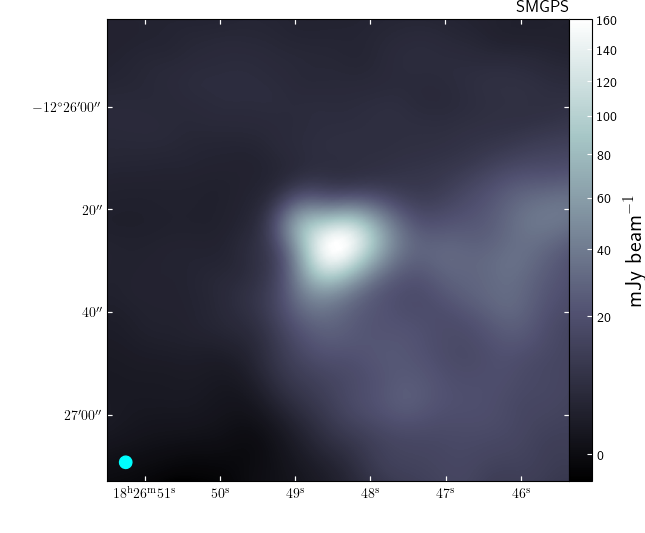}	
    \includegraphics[width=\columnwidth]{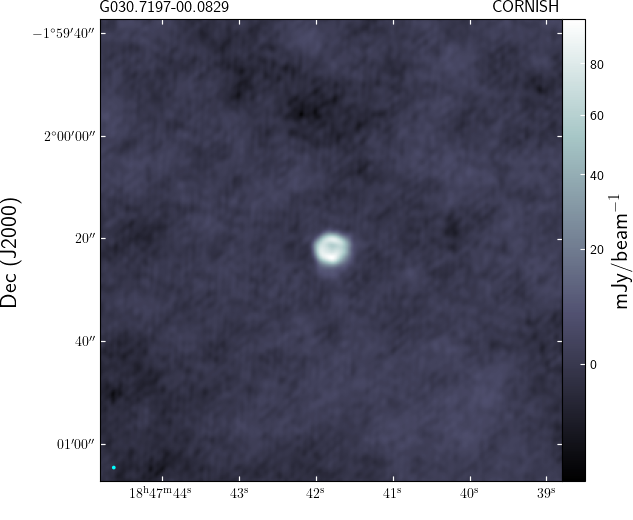}\includegraphics[width=\columnwidth]{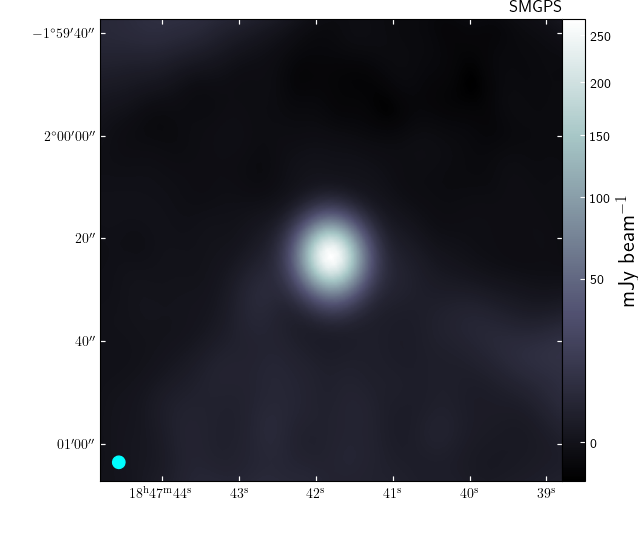}
    \includegraphics[width=\columnwidth]{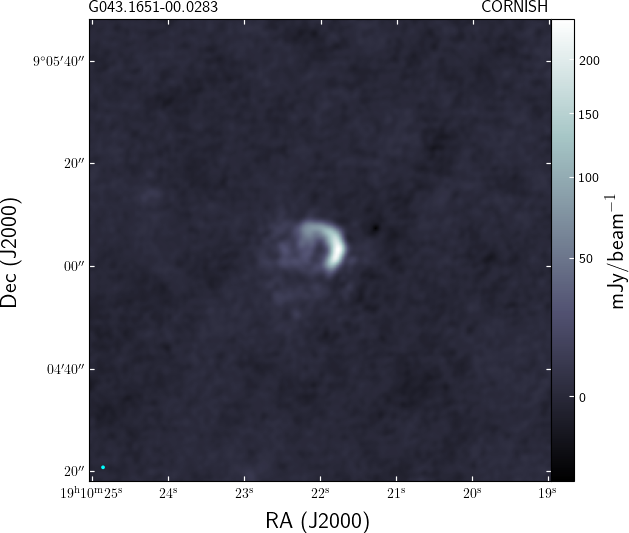}\includegraphics[width=\columnwidth]{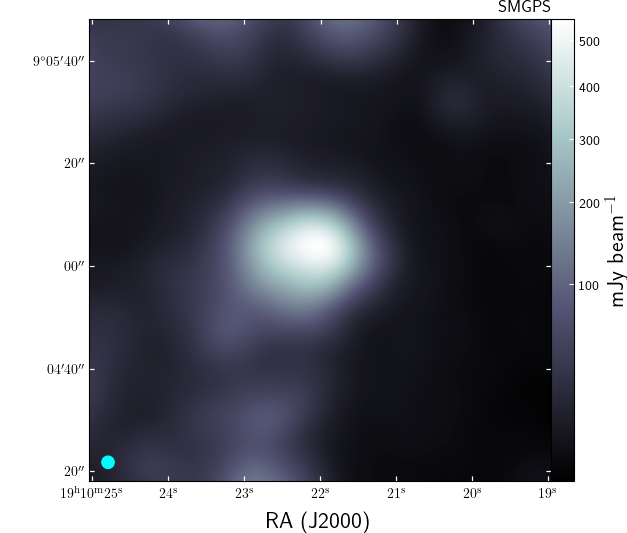}
    \caption{Example UC \hii\ regions selected from the CORNISH catalogue and plotted side-by-side with their SMGPS counterparts. The CORNISH images, in the left panels, show the higher resolution of the CORNISH survey (\ang[angle-symbol-over-decimal]{;;1.5}) and highlight the finer detail in the \hii\ regions, the SMGPS images in the right panels, imaged with the lower resolution of \ang{;;8}, are less resolved and show the diffuse emission surrounding the \hii\ regions. The filled circle in the lower left corner of the images represents their respective survey beams.}
    \label{fig:cornish_hiis}
\end{figure*}

The higher angular resolution in CORNISH reveals the finer structure in the \hii\ regions but filters out the more extended emission. The SMGPS, which has better \textit{uv}-coverage and is sensitive to the more diffuse emission, shows the extended structure that goes unseen in CORNISH. Fig.~\ref{fig:cornish_hiis} shows three side-by-side examples of \hii\ regions from CORNISH with their counterpart in the SMGPS. In the top row, CORNISH shows the finer detail in an UC \hii\ region appearing just outside the ionisation front. While the CORNISH source only shows the core of the UC \hii\ region and the ionisation front, the SMGPS also shows a tail of diffuse emission associated with the core. The diffuse emission surrounds both components portraying an unbroken source that shows a reduction in density of source emission farther from the peak. The middle row shows a source that is marginally unresolved in both surveys. However, here as well, in the SMGPS we see diffuse emission associated with the source. In the bottom row we see the resolved CORNISH source clearly showing the structure of a cometary UC \hii\ region that in the SMGPS is seen as a dense core of extended emission with a southward tail of diffuse emission. The differences in morphology seen in these comparisons highlight the hierarchical structure put forward by \cite{KimK2001} and also seen in \cite{Kurtz+1999} and \cite{Yang+2019}.

\subsection{The MMB-SMGPS association of methanol maser and continuum emission}
\cite{Walsh+1998} examined the association of methanol emission with radio continuum emission and showed how the projected sizes of the radio continuum regions are affected by the presence of methanol emission. They found that the radio continuum regions in which methanol emission is present are generally smaller than those in which it is not, which points to methanol emission being associated with the smallest, and therefore youngest, UC \hii\ regions. The implications of this result led to the hypotheses that the development of a 6.7 GHz methanol maser should  precede that of an UC \hii\ region, and the UC \hii\ region expansion eventually ``quenches'' the maser emission \citep{Walsh+1998}.  

 More recent studies have also examined the association of methanol masers with radio continuum emission. \cite{Hu+2016} conducted targeted VLA observations of 372 methanol masers and found 127 radio continuum counterparts associated with maser sources, giving a \s30 per cent rate of coincidence. \cite{Nguyen+2022} presented 554 maser detections from the GLOSTAR survey \citep{Brunthaler+2021} and found that 12 per cent of this maser population is coincident with radio continuum emission. The lower detection rate compared to \citet{Hu+2016} is attributed to the lower resolution and sensitivity of the GLOSTAR observations. Neither \citet{Hu+2016} or \citet{Nguyen+2022} found any correlation between maser luminosity and radio continuum luminosity.

\begin{figure}
    \includegraphics[width=\columnwidth]{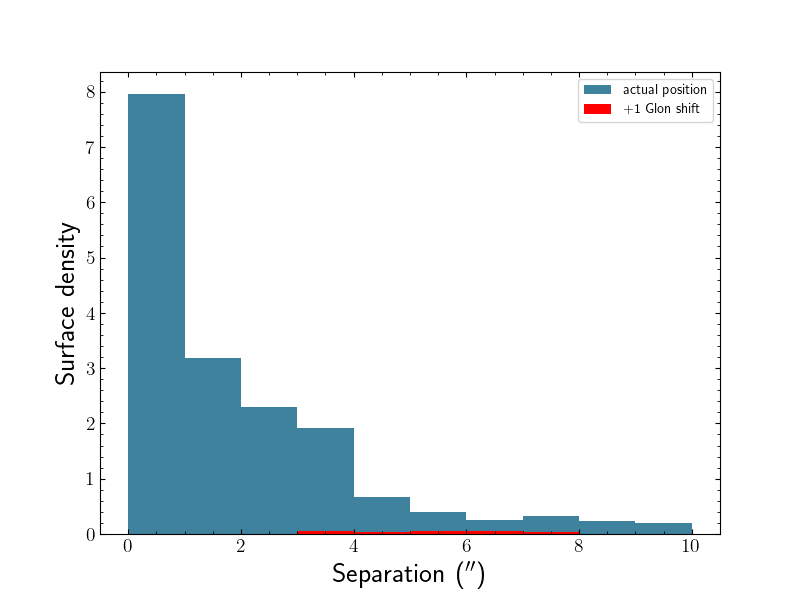}
    \caption{Surface density of the number of matches between the SMGPS and MMB as a function of angular separation. We show the result from crossmatching with the actual MMB positions in blue, and in red we show the distribution with the MMB positions shifted by 1\degr\ in Galactic longitude, to estimate the likelihood of chance alignments. The maximum separation was limited to $10''$, but the number of sources per bin is $< 10$ between $\sim$4\arcsec\ and 10\arcsec\ showing that while some of the masers in this range are coincident with an SMGPS source, most of them lie either on the boundary of, or far from their closest counterpart.}
    \label{fig:mmb-smgps_separation}
\end{figure}

The unprecedented wide area and sensitivity of the SMGPS offers the possibility to explore the correlation between 6.7 GHz methanol masers and radio continuum in a much larger sample ($\sim$ twice as large as the Nguyen sample) to greater continuum sensitivity than the Hu or Walsh targeted observations. We achieve this by cross-matching the SMGPS compact source catalogue to the Methanol Multi-Beam (MMB) Survey catalogue of 6.7 GHz masers \citep{Green+2009a, Green+2017}. Within the SMGPS survey area there are a total of 905 MMB 6.7 GHz masers, each of which are positioned to within an accuracy of  0.4 arcsec rms. We crossmatched these 905 masers with the 510\,599 compact radio sources from the SMGPS and show the resulting surface density plot in Fig.~\ref{fig:mmb-smgps_separation}. Taking a radius for true associations of 4\arcsec\ we find 140 MMB masers ($\sim$15 per cent of the total population) to be associated with SMGPS compact continuum sources. We estimate the likelihood of chance alignments by shifting the positions of the MMB masers by 1\degr\ in Galactic longitude and repeating the crossmatch. We show the resulting surface density plot in the red histogram of Fig.~\ref{fig:mmb-smgps_separation}. The crossmatch between the SMGPS and the longitude-shifted MMB catalogue resulted in 9 matches within 10\arcsec, only one of which has separation < 4\arcsec. Therefore, we estimate the likelihood of chance alignments to be \s0.7 per cent.

The MMB-SMGPS detection rate is consistent with that of \citet{Hu+2016}, reinforcing the conclusions of \citet{Nguyen+2022} that their different detection rate is due to different continuum sensitivity. This detection rate also confirms the original result from \citet{Walsh+1998} that the majority of methanol maser emission are not associated with radio continuum emission. Similarly, most sites of radio continuum emission show no evidence of methanol maser emission. We examined the relationship between maser luminosity and 1.3 GHz continuum luminosity and found no significant correlation between the two quantities, again confirming the results of \citet{Hu+2016,Nguyen+2022}.

While higher angular resolutions are required to resolve the SMGPS sources at the sub-arcsecond scale of UC \hii\ regions, this study supports the \cite{Walsh+1998} hypothesis that methanol masers are more likely to be seen around an UC \hii\ region when the \hii\ region is small and young, before it destroys the maser \citep{Walsh+1997}.

\section{Conclusions}
\label{sec:conclusions}
We present a catalogue of compact (< 5 synthesised beam areas) sources detected in the 1.3 GHz SARAO MeerKAT Galactic Plane Survey. The final catalogue has 510\,599 distinct components detected with a signal-to-noise ratio of 5 or greater. These predominantly comprise isolated components but include sources that are members of larger complexes resolved into multiple components. The catalogue has a total of 489\,542 source islands, with 473\,421 ($\sim$97 per cent) being detected as single components and 16\,121 as multiple components: 13\,206 two-component islands, 2054 three-component islands, and 861 islands with four or more components.

We assess the fidelity of the catalogue through completeness checks and false positive rate estimates. The complex environment of the Galactic Plane resulted in uneven detection of sources across the survey. However, across the representative tiles, we determine better than 90 per cent completeness for sources with S/N $\geq5$. We examine the effect of background noise on the detected flux densities by assessing the extent of flux boosting in the catalogue sources. We find the deviation in flux density to be between 2 and 9 per cent at the lowest flux densities, with the deviation only going up to 9 per cent at $S_{\text{peak}} \leq$ 0.1 \mjy\, essentially showing a 1-to-1 recovery for the majority of the sources.

The rate of false positives is estimated to be 0.63 per cent. The highest percentage of false positives occur just below 5$\sigma$. Therefore, following these checks, we choose to only include sources in our catalogue with an S/N $\geq$ 5, and as the effect of flux boosting is less than 10 per cent at low flux densities, we also choose not to correct the SMGPS compact source catalogue for flux boosting. 

We examine the morphology of the compact sources through the source elongations ($a/b$) and the 1.3 GHz $Y$-factor ($S_{\text{integrated}}/S_{\text{peak}}$). The majority of the sources show low elongation ratios with a median elongation ratio of 1.1, indicating a predominantly spherical or low axis-ratio population. In examining the $Y$-factor, which generally should equal 1 for point or unresolved sources, we find that the majority of the compact sources in the catalogue are marginally resolved and peak at a $Y$-factor of $\sim$1.2.

Through a cross-reference with the SIMBAD database, we explore the nature of the sources in the SMGPS compact source catalogue. The majority of the bright sources in the catalogue are contained within the literature; \s62 per cent of the sources with flux densities above 100 mJy are associated with SIMBAD objects. However, this population of previously seen sources represents only $\sim$1 per cent of the survey. The bulk of the unknown population are at sub-mJy levels, underlining the importance of such a deep survey.

We investigate the radio quiet population of the WISE \hii\ region catalogue for associations with compact radio continuum emission from the SMGPS. We detected compact radio continuum emission associated with 213 WISE \hii\ region candidates classified as radio quiet. These sources are likely to have gone unseen in previous radio surveys because they predominantly have fainter radio flux densities. However, determining their true nature is left to future studies, as the multi-wavelength analysis required for this is beyond the scope of this paper. 

A comparison of SMGPS compact sources with CORNISH UC \hii\ regions shows that the higher angular resolution in CORNISH reveals the finer structure in the \hii\ regions while the SMGPS, which has better \textit{uv}-coverage and is sensitive to more diffuse emission, shows the extended emission not seen in CORNISH.

Finally, we examine the association of methanol emission with radio continuum emission by comparison with the methanol multibeam survey. We determine 140 matches in a pool of 905 masers (\s15 per cent) thus supporting the \cite{Walsh+1998} hypothesis: methanol masers are more likely to be associated with small, young \hii\ regions than the larger, more evolved \hii\ regions because an expanding \hii\ region is likely to overrun and disrupt the maser.

The survey depth and sensitivity has resulted in a population of previously unknown detections at low flux density thus bringing new perspective and presenting great potential for synergy with other surveys, both in the radio and other wavelengths.

\section*{Acknowledgements}

We thank the referee for an insightful and constructive report that helped improve the quality of this paper. The MeerKAT telescope is operated by the South African Radio Astronomy Observatory (SARAO), which is a facility of the National Research Foundation, an agency of the Department of Science and Innovation. MM, MAT, MGH, and GMW gratefully acknowledge the support of the UK’s Science \& Technology Facilities Council (STFC) through grant awards ST/T000287/1 and ST/X001016/1. This research has made use of NASA's Astrophysics Data System Bibliographic Services, the radio source finder {\sc Aegean} \citep{Hancock+2012}, the Starlink Tables Infrastructure Library Tool Set \citep[{\sc stilts}:][]{stilts}, and {\sc python} packages {\sc astropy} \citep{astropy}, {\sc matplotlib} \citep{matplotlib}, {\sc numpy} \citep{numpy}, and {\sc scipy} \citep{scipy}.

\section*{Data Availability}

 The full SMGPS compact source catalogue is made available at: \url{https://doi.org/10.48479/h16d-2z55}. The SMGPS Data Release 1 (DR1; \citealt{Goedhart+2024}) containing the images from which the catalogue was extracted is available at:  \url{https://doi.org/10.48479/3wfd-e270}.


\bibliographystyle{mnras}
\bibliography{mgps_psc_bibliography} 



\appendix

\section{Catalogue Format}
We present a catalogue of compact sources. While an excerpt of the catalogue with a subset of columns is shown in Table~\ref{tab:smgps_catalogue_snippet}, here we present the format of the columns in the full catalogue.

\clearpage
\begin{table}
\caption{Format of the SMGPS catalogue of compact sources}
\label{tab:smgps_catalogue_format}
\begin{tabular}{clcl}
\hline
Col. Num. & Name                     & Unit            & Description                                                                         \\
\hline
1         & csc\_id        & -               & Catalogue source identification number                                                  \\
2         & Name          & -               & Galactic source name                                                                    \\
3         & IAUName       & -               & IAU designation of the source name                                                      \\
4         & tileName      & -               & File identifier of mosaic from which the source was extracted                           \\
5         & island        & -               & A group of 1 or more Gaussians identified as pixels with contiguous source emission     \\
6         & source        & -               & A single Gaussian component                                                             \\
7         & background    & mJy beam$^{-1}$ & background flux density                                                                 \\
8         & local\_rms     & mJy beam$^{-1}$ & rms noise around the immediate vicinity of the source                                   \\
9         & lon           & deg             & Galactic longitude of the centroid position of the source                               \\
10        & err\_lon       & deg             & Source-extraction fitting error on Galactic longitude                                   \\
11        & lat           & deg             & Galactic latitude of the centroid position of the source                                \\
12        & err\_lat       & deg             & Source-extraction fitting error on Galactic latitude                                    \\
13        & ra            & deg             & Right Ascension of the centroid position of the source                                  \\
14        & dec           & deg             & Declination of the centroid position of the source                                      \\
15        & peak\_flux     & mJy beam$^{-1}$ & Peak flux density                                                                       \\
16        & err\_peak\_flux & mJy beam$^{-1}$ & Source-extraction fitting error on peak flux density                                    \\
17        & int\_flux      & mJy             & Integrated flux density (calculated as a/b/peak\_flux)                                   \\
18        & err\_int\_flux  & mJy             & Source-extraction fitting error on integrated flux density                              \\
19        & a             & arcsec          & Fitted major axis                                                                       \\
20        & err\_a         & arcsec          & Error on fitted major axis                                                              \\
21        & b             & arcsec          & Fitted minor axis                                                                       \\
22        & err\_b         & arcsec          & Error on fitted minor axis                                                              \\
23        & pa            & deg             & Fitted position angle                                                                   \\
24        & err\_pa        & deg             & Error on fitted position angle                                                          \\
25        & flags         & -               & Fitting flags (see main text for description)                                           \\
26        & snr           & -               & Signal-to-noise ratio (calculated as peak\_flux/local\_rms)                               \\
27        & area          & arcsec$^2$      & Source area (calculated as $\frac{\pi \cdot \theta_{maj} \cdot \theta_{min}}{4\ln{2}}$) \\
28        & nbs           & -               & Near bright source flag (yes = within 0.5\degr\ of a bright source)                     \\
\hline
\end{tabular}
\end{table}


\bsp	
\label{lastpage}
\end{document}